\documentclass[twocolumn,showpacs,preprintnumbers,amsmath,aps,longbibliography]{revtex4-1}
\usepackage{graphicx}
\usepackage{dcolumn}
\usepackage{bm}
\usepackage{multirow}
\usepackage{tabularx}
\usepackage{color}
\usepackage{stmaryrd}
\begin{document}
%%%%%%%%%%%%%%%%%%%%%%%%%%%%
\def\eq#1{Eq.(\ref{#1})}
\def\fig#1{Fig.\hspace{1mm}\ref{#1}}
\def\tab#1{Table\hspace{1mm}\ref{#1}}
%%%%%%%%%%%%%%%%%%%%%%%%%%%%
\title{
%---------------------------------------------------------------------------------------------------------------\\
Strong-coupling superconductivity induced by calcium intercalation in bilayer transition-metal dichalcogenides
}

\author{R. Szcz{\c{e}}{\'s}niak}
\author{A.P. Durajski}\email{adurajski@wip.pcz.pl}
\author{M.W. Jarosik}
%%%%%%%%%%%%
\affiliation{Institute of Physics, Cz{\c{e}}stochowa University of Technology, Ave. Armii Krajowej 19, 42-200 Cz{\c{e}}stochowa, Poland}
%%%%%%%%%%%%
\begin{abstract}
%%%%%%%%%%%%%%%%%%%%%%%%%%%%%%%%%%%%%%%%%%%%%
%
We theoretically investigate the possibility of achieving a superconducting state in transition-metal dichalcogenide bilayers through intercalation, a process previously and widely used to achieve metallization and superconducting states in novel superconductors.
For the Ca-intercalated bilayers MoS$_2$ and WS$_2$, we find that the superconducting state is characterized by an electron-phonon coupling constant larger than $1.0$ and a superconducting critical temperature of $13.3$ and $9.3$ K, respectively.
These results are superior to other predicted or experimentally observed two-dimensional conventional superconductors and suggest that the investigated materials may be good candidates for nanoscale superconductors.
More interestingly, we proved that the obtained thermodynamic properties go beyond the predictions of the mean-field Bardeen--Cooper--Schrieffer approximation and that the calculations conducted within the framework of the strong-coupling Eliashberg theory should be treated as those that yield quantitative results.
\\\\
Keywords: 2D superconductivity, Effect of intercalation, Transition-metal dichalcogenides, Thermodynamic properties
%%%%%%%%%%%%%%%%%%%%%%%%%%%%%%%%%%%%%%%%%%%%%
\end{abstract}
\vspace{-0.5cm}
\pacs{74.20.Fg, 74.25.Bt, 74.62.Fj}
\maketitle

%
%%%%%%%%%%%%%%%%%%%%%%%%%%%%%%%%%%%%%%%%%%%%%%%%%%%%%%%%%%%%%%%%%%%%%%%%%%%%%%%%%%%%%%%%%%%%%%%%%%%
\section{Introduction}
%%%%%%%%%%%%%%%%%%%%%%%%%%%%%%%%%%%%%%%%%%%%%%%%%%%%%%%%%%%%%%%%%%%%%%%%%%%%%%%%%%%%%%%%%%%%%%%%%%%

Superconductivity in two-dimensional (2D) materials has attracted considerable interest since it was noted that 2D monolayers may exhibit different properties than their corresponding bulk materials \cite{Saito, WonbongChoi, Uchihashi}. 2D superconductivity has been investigated for the past $80$ years, and the research has provided insight into a variety of quantum phenomena such as localization of electrons and/or Cooper pairs \cite{Sacepe}, oscillations of order parameter and critical temperature caused by quantum size effects \cite{GuoYang, DurajskiPb, Talantsev}, transition from a superconducting to an insulating phase with increasing disorder or magnetic field \cite{GoldmanMarkovic} and Berezinskii--Kosterlitz--Thouless transition \cite{Berezinskii1, Berezinskii2, Berezinskii3}.
Recent advances in these materials have led to a variety of promising technologies for nanoelectronics, photonics, sensing, energy storage, and optoelectronics \cite{Zhang2017}.

Over the past several years, special attention has been paid on transition-metal dichalcogenides ($MX_2$, where $M$ = Mo or W, and $X$=S, Se, or Te) because of their fascinating physical properties and their potential for various applications. Particularly, molybdenum disulfide (MoS$_2$) has attracted significant interest \cite{WonbongChoi, HuaWang, DuanXidong, YuzhengGuo, DominMoS, AliKandemir, PujuZhao}. Bulk MoS$_2$ is a semiconductor with an indirect band gap of about $1.29$ eV \cite{KamMoS2}, while its monolayer form has a direct band gap of $1.90$ eV \cite{MakMoS2}. This means that MoS$_2$ is a prime candidate for use in optoelectronic devices, transistors, and photodetectors \cite{HuaWang, Luo2017}.
Similar to graphene \cite{novoselov, JiangMoS2Graphene, DurajskiGraphene}, semiconducting layered transition-metal dichalcogenides can be metallized or even become superconductors upon alkali metal-atom intercalation or strain \cite{XinyueLin, JelenaPesic, ZhangMoS}.
Theoretical research has shown that Li- and Na-intercalated bilayer MoS$_2$ are promising conventional superconductors having superconducting transition temperatures of approximately $10.2$ K and $2.8$ K, respectively \cite{HuangMoS, ZhangMoS}.
More interestingly, $T_C$ and electronic properties can be significantly enhanced by tensile strain \cite{XinHe}. In the case of (MoS$_2$)$_2$Na at $7\%$, biaxial tensile strain $T_C$ increases to $10$~K~\cite{ZhangMoS}.
Experimentally, superconductivity in the MoS$_2$ transistor that adopts an electric double layer as a gate dielectric was successfully induced with a maximum $T_C$ of approximately $10.8$ K \cite{Ye1193}.
In addition, after external pressure was applied, iso-structural semiconducting to metallic transition in multilayered MoS$_2$ was observed at $\sim$19 GPa \cite{Nayak}, and the emergence of a superconducting state having the highest critical temperature among transition metal dichalcogenides ($T_C$ of approximately $11.5$ K above $120$ GPa) was reported \cite{ZhenhuaChi}.
Unfortunately, from the technological point of view, such high pressure excludes MoS$_2$ from potential practical applications. Because of this, finding intercalants that, after being introduced into layered structures, allow for induction of a superconducting state with as high a critical temperature at normal pressure as possible \cite{Somoano, Somoano2, Somoano3, GQHuang, YuSaito, DominCLi} seems natural.

In the present study, we combine first-principles density functional theory and strong-coupling Eliashberg formalism for a comparative study of superconductivity in Ca-intercalated transition-metal dichalcogenide bilayers MoS$_2$ and WS$_2$. 
The choice of Ca as an intercalant was inspired by recent studies of experimentally observed superconductivity in Ca-intercalated bilayer graphene and Ca-doped graphene laminates with $T_C$ of approximately $4$ K and $6.4$ K, respectively \cite{IchinokuraCaC1, ChapmanCaC1}.
Our results show that (MoS$_2$)$_2$Ca and (WS$_2$)$_2$Ca systems are phonon-mediated strong-coupling superconductors having critical temperatures of $13.3$ and $9.3$ K, respectively.

%
%%%%%%%%%%%%%%%%%%%%%%%%%%%%%%%%%%%%%%%%%%%%%%%%%%%%%%%%%%%%%%%%%%%%%%%%%%%%%%%%%%%%%%%%%%%%%%%%%%%
\section{Computational details}
\begin{figure} [!t]       
\includegraphics[width=0.9\columnwidth]{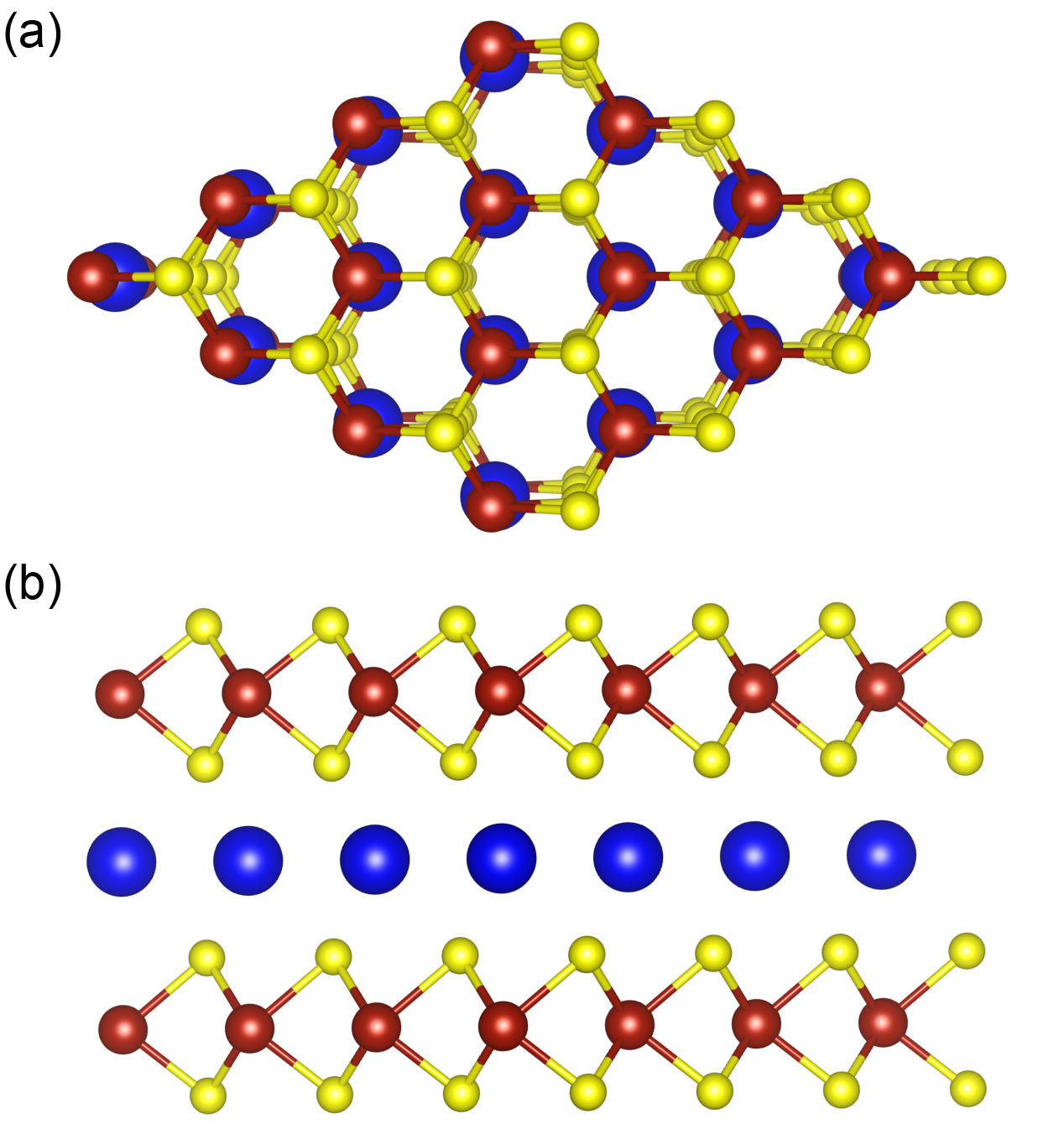} 
\caption{Top (a) and front (b) view of the ($M$S$_2$)$_2$Ca superconductor. The red, yellow, and blue spheres denote Mo (or W), S, and Ca atoms, respectively.
In order to avoid any interaction between intercalated bilayers, we used the periodic boundary condition with a vacuum space of $20~\rm \AA$ along the non-periodic $z$ direction.}         
\label{f0}
\end{figure}

\begin{figure} [!t]       
\includegraphics[width=1\columnwidth]{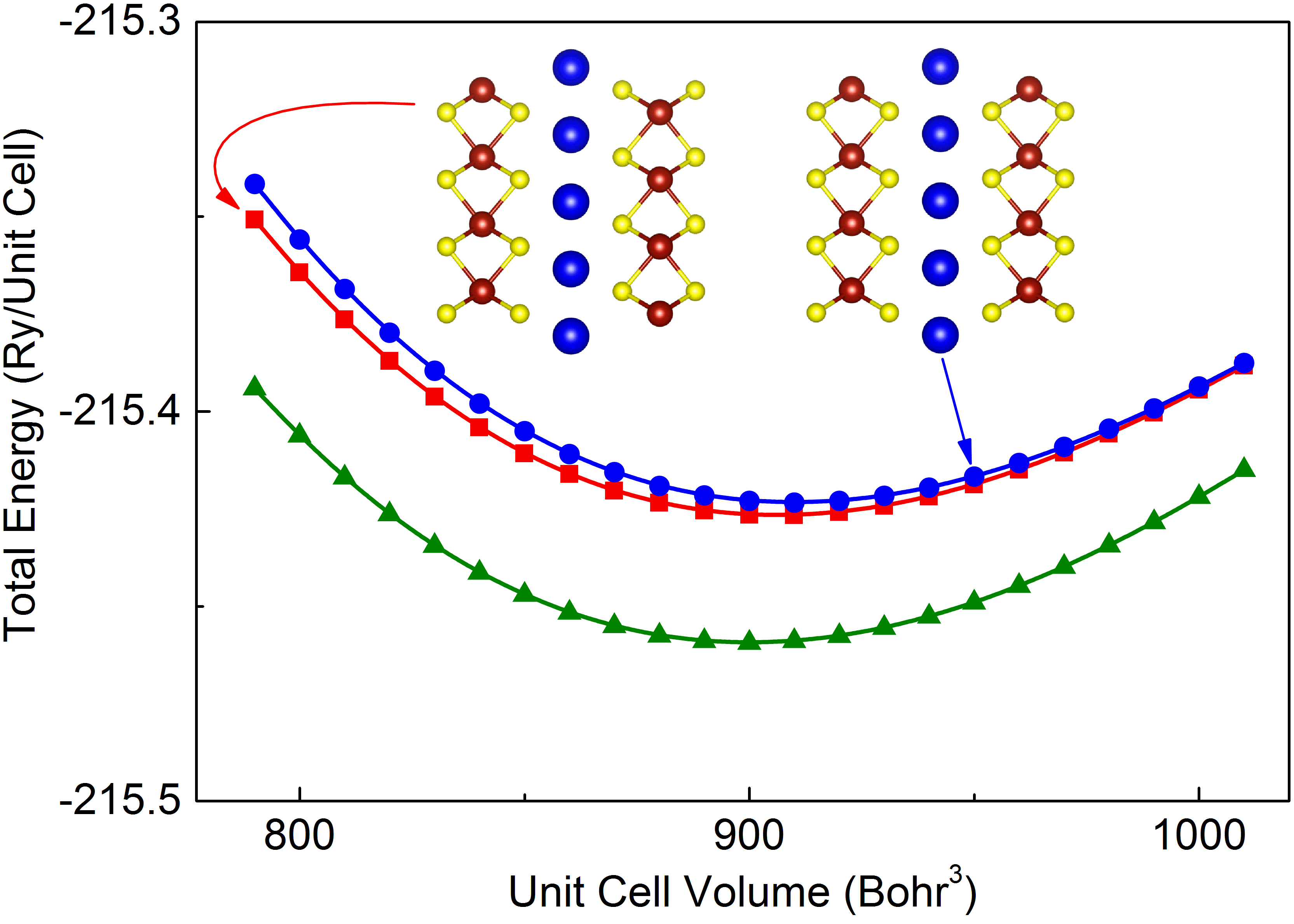} 
\caption{The calculated total energy vs. volumes for three possible stacking orders of Ca-intercalated bilayer MoS$_2$ (the green line corresponds to the structure presented in \fig{f0}).}
\label{f0a}
\end{figure}

\begin{figure}        
\includegraphics[width=\columnwidth]{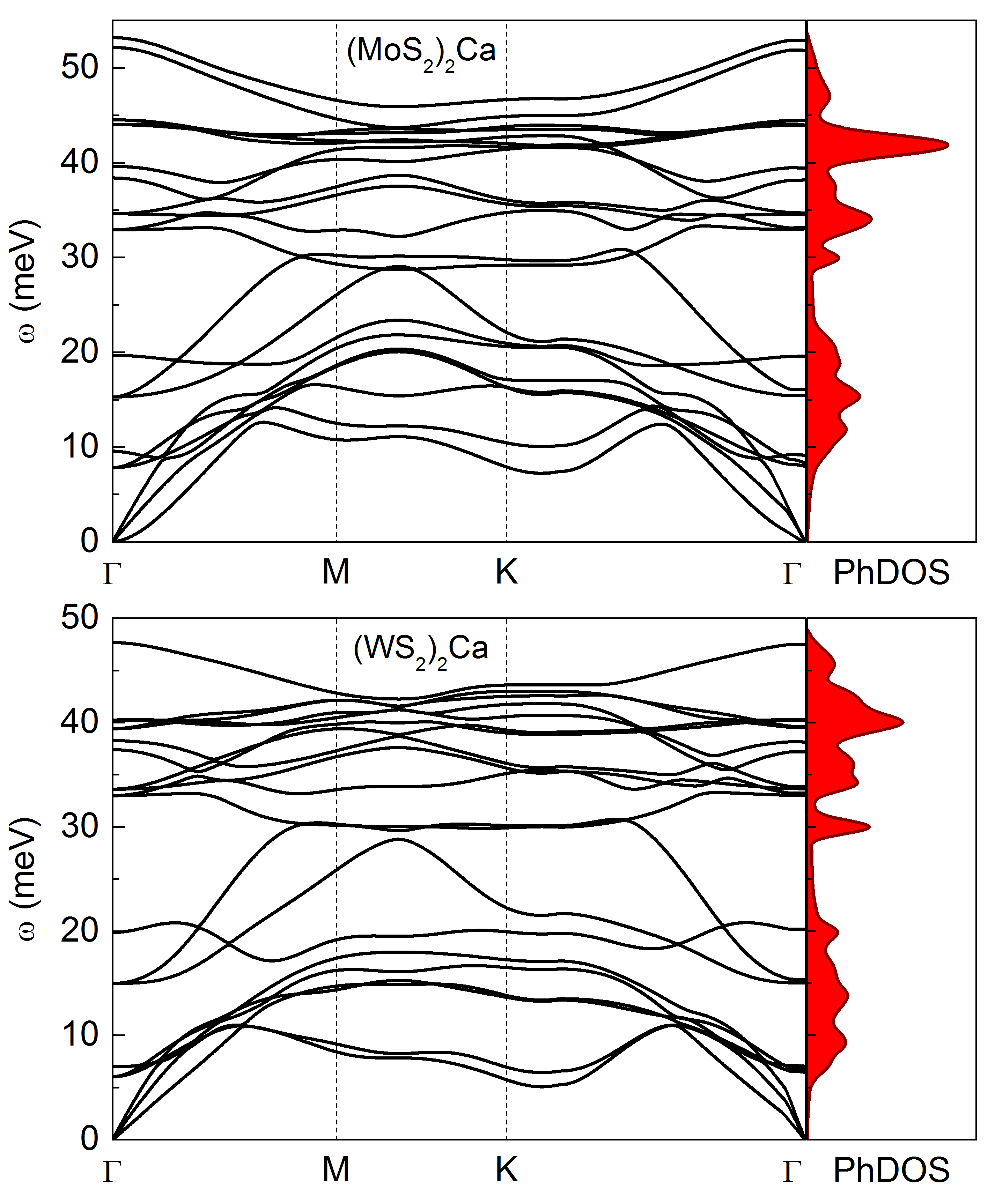} 
\caption{Phonon dispersion and phonon density of states for Ca-intercalated bilayer MoS$_2$ and WS$_2$.}         
         \label{f1}
\end{figure}

%%%%%%%%%%%%%%%%%%%%%%%%%%%%%%%%%%%%%%%%%%%%%%%%%%%%%%%%%%%%%%%%%%%%%%%%%%%%%%%%%%%%%%%%%%%%%%%%%%%
First-principles studies are performed within the framework of the density functional theory (DFT) as implemented in the Quantum-ESPRESSO package \cite{Giannozzi2009A, GiustinoEP, FelicianoGiustino}. 
The generalized gradient approximation (GGA) with the Perdew--Wang (PW91) exchange correlation function was used in our study. The plane-wave energy cutoff for the wavefunctions was set to $80$ Ry. For the electronic structure investigations, the Brillouin zone was sampled using a set of $\rm 24\times 24\times 1$ Monkhorst--Pack $k$-mesh. 
A $k$-points grid of $\rm 60\times 60\times 1$ and a $q$-points grid of $\rm 6\times 6\times 1$ were used to calculate the electron-phonon coupling matrix and phonon spectrum. 
The single-layer of transition-metal dichalcogenide consists of one monoatomic hexagonal Mo or W plane placed between two monoatomic hexagonal S planes that are bonded together through weak van der Waals interactions \cite{vanderWalls}. Following intercalation, two transition-metal dichalcogenide layers were separated by foreign-atoms layer. 
By performing structural relaxation, we confirmed that for the most stable structure of bilayers MoS$_2$ and WS$_2$ intercalated by Ca atoms, the top and bottom monolayers are mirror images of each other and the intercalated Ca atoms have the same configuration as that of the Mo or W atoms (see \fig{f0}). 
The previously conducted first-principles calculations show that, from among three possible stacking orders, this structure is dynamically stable for Li-intercalated bilayer MoS$_2$ \cite{HuangMoS}.
The total energy of (MoS$_2$)$_2$Ca as a function of the volumes for three different structural types are given in \fig{f0a}. We can see that the structure from \fig{f0} has lower total energy at an equilibrium volume.
The presence of imaginary phonon frequencies indicates structural instability. The calculated phonon dispersion and phonon density of states (PhDOS) of Ca-intercalated bilayer molybdenum and tungsten disulfides are shown in \fig{f1}. We can see that no negative frequencies exist, thus confirming the dynamical stability of the structure from \fig{f0}. Therefore, we use this structure to calculate the electron-phonon coupling coefficients and thermodynamic properties of (MoS$_2$)$_2$Ca and (WS$_2$)$_2$Ca in the superconducting state.
At this point, we should mention that in the work of Somoano \textit{et al.} \cite{Somoano}, bulk Ca-intercalated molybdenum disulfide crystal was synthesized, characterized, and measured. As the authors found, Ca-intercalated MoS$_2$ crystallizes in an orthorhombic structure, which is in contrast to the hexagonal structure of pristine MoS$_2$ and alkali-metal-intercalated MoS$_2$. The origin of this difference between bilayer and bulk structure is unknown; additional research is required.

\begin{figure}[!t]       
\includegraphics[width=1\columnwidth]{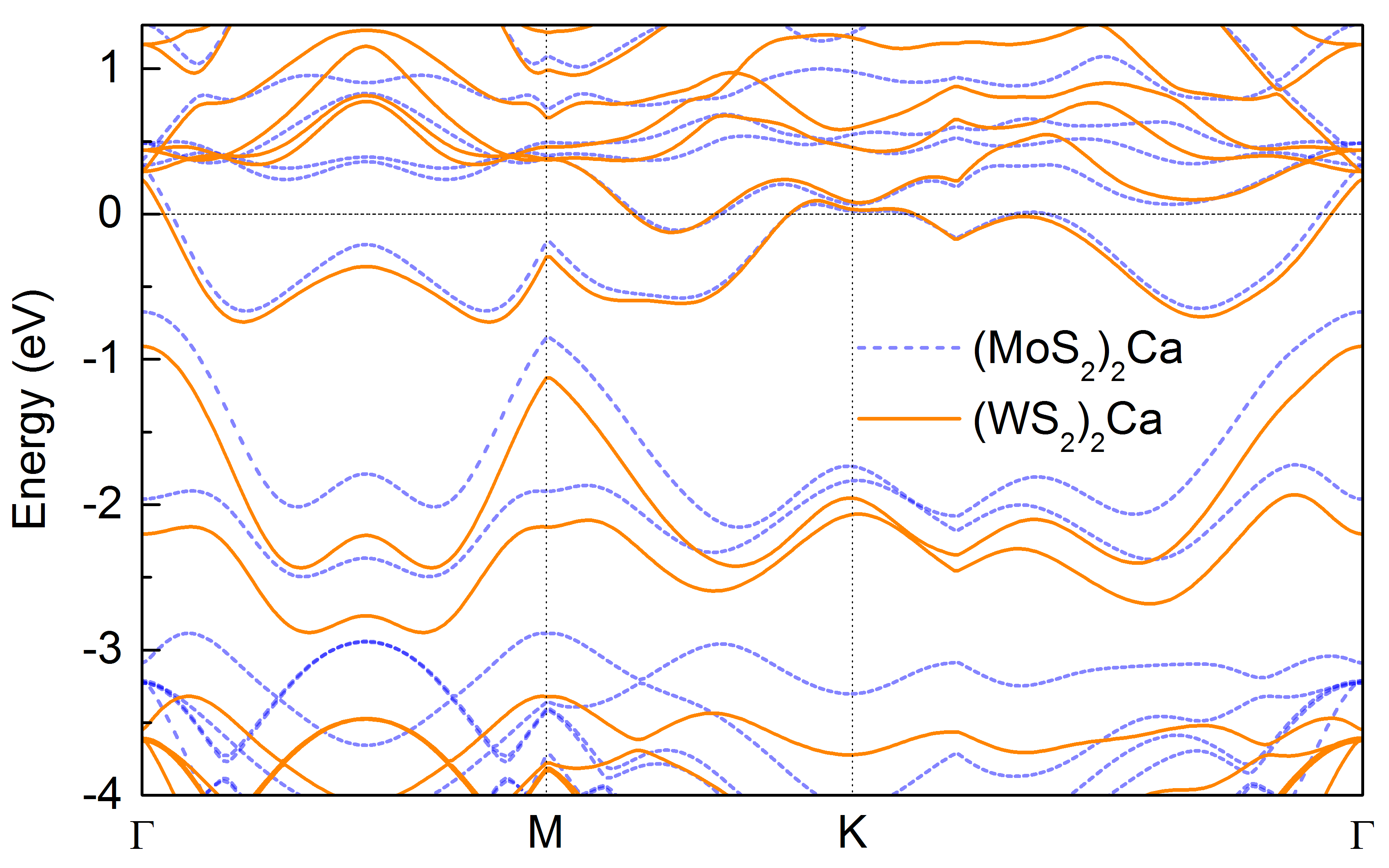} 
\caption{Calculated electronic band structures of (MoS$_2$)$_2$Ca (dashed blue lines) and (WS$_2$)$_2$Ca (solid orange lines).}
\label{f1a}
\end{figure}

Before calculating possible superconducting properties, we analyzed the electronic band structures of (MoS$_2$)$_2$Ca and (WS$_2$)$_2$Ca. In \fig{f1a}, the band dispersion shows the metallic character of both systems. This means that after Ca-atoms intercalation of transition-metal dichalcogenides, we can observe the transition from semiconducting to the metallic phase \cite{SzczDurJar}.

\begin{figure}[!t]
\includegraphics[width=\columnwidth]{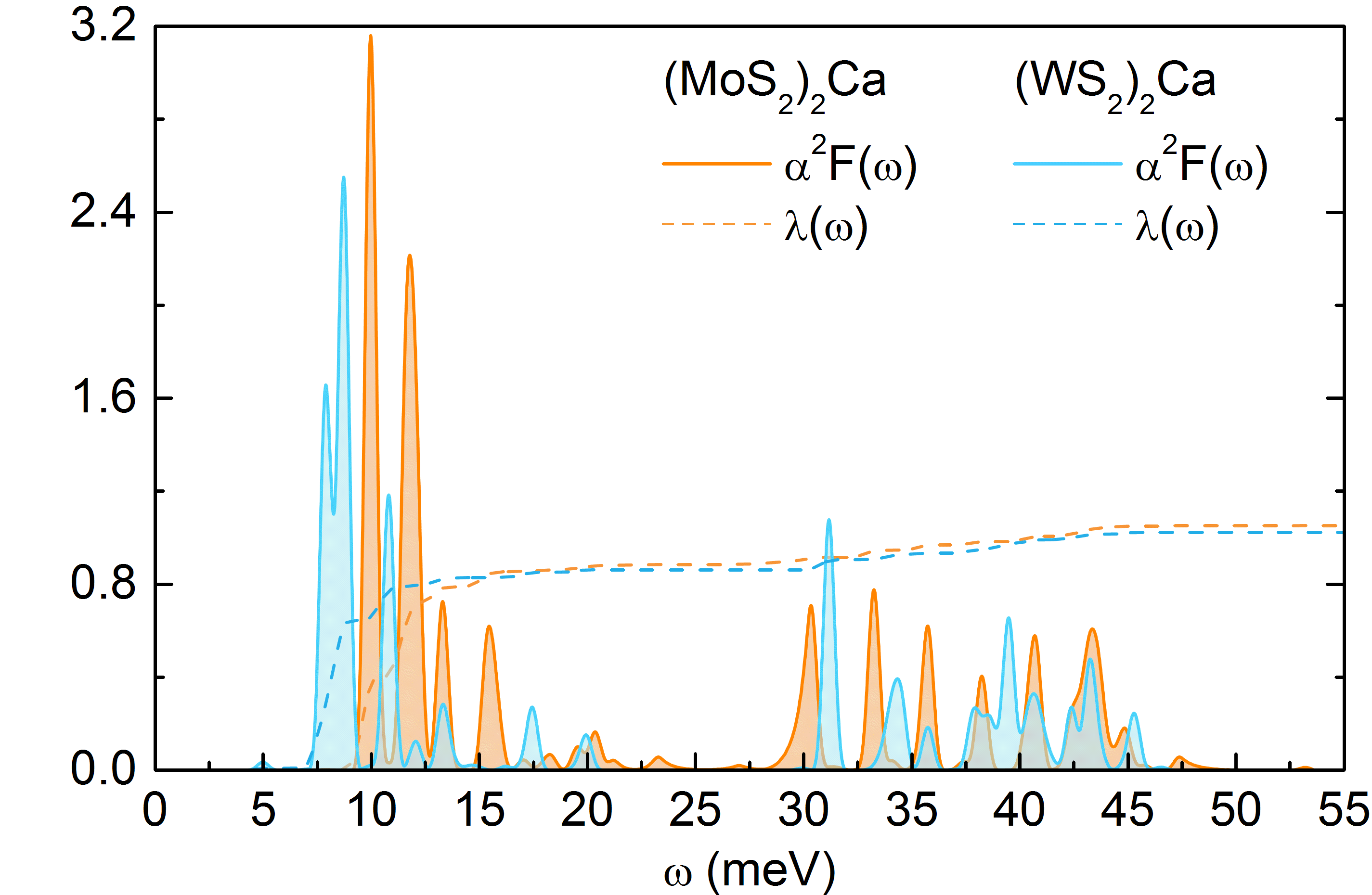} 
\caption{Eliashberg spectral function $\alpha^2F(\omega)$ and the cumulative electron-phonon coupling strength $\lambda(\omega)$ of Ca-intercalated bilayer MoS$_2$ and WS$_2$.}
\label{f2}
\end{figure}

To determine the thermodynamic properties of a conventional superconductor, the standard approach is to solve numerically the Eliashberg equations \cite{Eliashberg1960A, Carbotte1990A, Szczesniak2006B, Szczesniak2016MgH6}:
\begin{equation}
\label{r1}
\Delta_{m}Z_{m}={\pi}{k_BT}\sum_{n}
\frac{\lambda_{n,m}-\mu^{\star}\theta\left(\omega_{c}-|\omega_{n}|\right)}
{\sqrt{\omega_n^2Z^{2}_{n}+\varphi^{2}_{n}}}\varphi_{n},
\end{equation}
and
\begin{equation}
\label{r2}
Z_{m}=1+\frac{\pi k_BT}{\omega_{n}}\sum_{n}
\frac{\lambda_{n,m}}{\sqrt{\omega_n^2Z^{2}_{n}+\varphi^{2}_{n}}}
\omega_{n}Z_{n},
\end{equation}
where the first equation is for the superconducting order parameter function and the second determines the electron effective mass.
The pairing kernel for the electron-phonon interaction is given by:
\begin{equation}
\label{r3}
\lambda_{n,m}= 2\int_0^{\omega_{{D}}}d\omega\frac{\omega}
{\left(\omega_n-\omega_m\right)^2+\omega ^2}\alpha^{2}F\left(\omega\right).
\end{equation}
$k_B$, $\mu^{\star}$, and $\theta$ denote the Boltzmann constant ($0.0862$ meV/K), the Coulomb pseudopotential (we use the commonly accepted value of $0.1$), and the Heaviside step function with cut-off frequency $\omega_c$ equal to three times the maximum phonon frequency $\omega_D$.
The Eliashberg spectral function $\alpha^2F(\omega)$ is given by:
\begin{equation}
  \alpha^2F(\omega)=\frac{1}{2\pi N(\epsilon_F)}\sum_{\mathbf{q}v}
  \delta(\omega-\omega_{\mathbf{q}v})\frac{\gamma_{\mathbf{q}v}}{\hbar \omega_{\mathbf{q}v}},
\end{equation}
where the line width of phonon mode $v$ at the wave vector $\mathbf{q}$ is defined as follows:
\begin{equation}
\begin{split}
\gamma_{\mathbf{q} v}&=2\pi \omega_{\mathbf{q} v}\sum_{ij}\int\frac{\mathrm{d^3}{k}}{\Omega_{BZ}}
|g_{\mathbf{q}v}(\mathbf{k},i,j)|^2 \delta(\epsilon_{\mathbf{q},i}-\epsilon_F)\\ &\times \delta(\epsilon_{\mathbf{k+q},j}-\epsilon_F) ,
\end{split}
\end{equation}
Symbol $\Omega_{BZ}$ refers to the value of the Brillouin zone, $i$ and $j$ are band indices, $\epsilon_{\mathbf{q},i}$ denotes the energy of the bare electronic Bloch state, and $g_{\mathbf{q}v}(\mathbf{k},i,j)$ is the electron-phonon matrix element that can be determined in a self-consistent manner by the linear response theory.
The shape of the Eliashberg function for Ca-intercalated bilayer molybdenum and tungsten disulfides are shown in \fig{f2}.
In addition, the frequency-dependent electron-phonon coupling, given by $\lambda(\omega)=2\int_0^{\omega_{{D}}}{\omega^{-1}}{\alpha^2F(\omega)}d\omega$, is plotted using dashed lines. The obtained $\lambda(\omega_D)=1.05$ for (MoS$_2$)$_2$Ca and $\lambda(\omega_D)=1.02$ for (WS$_2$)$_2$Ca clearly indicates that we are dealing with strong-coupling superconductors.

The numerical solutions of \ref{r1} and \ref{r2} enable us to determine the condensation energy ($E_{\rm cond}$) defined as a difference between the free energy of the normal state and that of the superconducting state in the absence of a magnetic field \cite{Carbotte1990A, CARBOTTE1970227}:
\begin{eqnarray}
\label{rF}
\frac{E_{\rm cond}(T)}{N(\epsilon_F)}&=&-\pi T\sum_{n}
\left(\sqrt{\omega^{2}_{n}+\Delta^{2}_{n}}- \left|\omega_{n}\right|\right)\\ \nonumber
&\times&\left(Z^{N}_{n}\frac{\left|\omega_{n}\right|}
{\sqrt{\omega^{2}_{n}+\Delta^{2}_{n}}}-Z^{S}_{n}\right),
\end{eqnarray}  
where $Z^{N}_{n}$ and $Z^{S}_{n}$ denote the mass renormalization functions for the normal ($\Delta_{m}=0$) and superconducting ($\Delta_{m}>0$) states, respectively.
In this study, the calculations of thermodynamics of phonon-mediated superconductors (MoS$_2$)$_2$Ca and (WS$_2$)$_2$Ca are based mostly on the results obtained for the condensation energy.
%%%%%%%%%%%%%%%%%%%%%%%%%%%%%%%%%%%%%%%%%%%%%%%%%%%%%%%%%%%%%%%%%%%%%%%%%%%%%%%%%%%%%%%%%%%%%%%%%%%
\section{Results and Discussion}
\begin{figure} [th]        
\includegraphics[width=1\columnwidth]{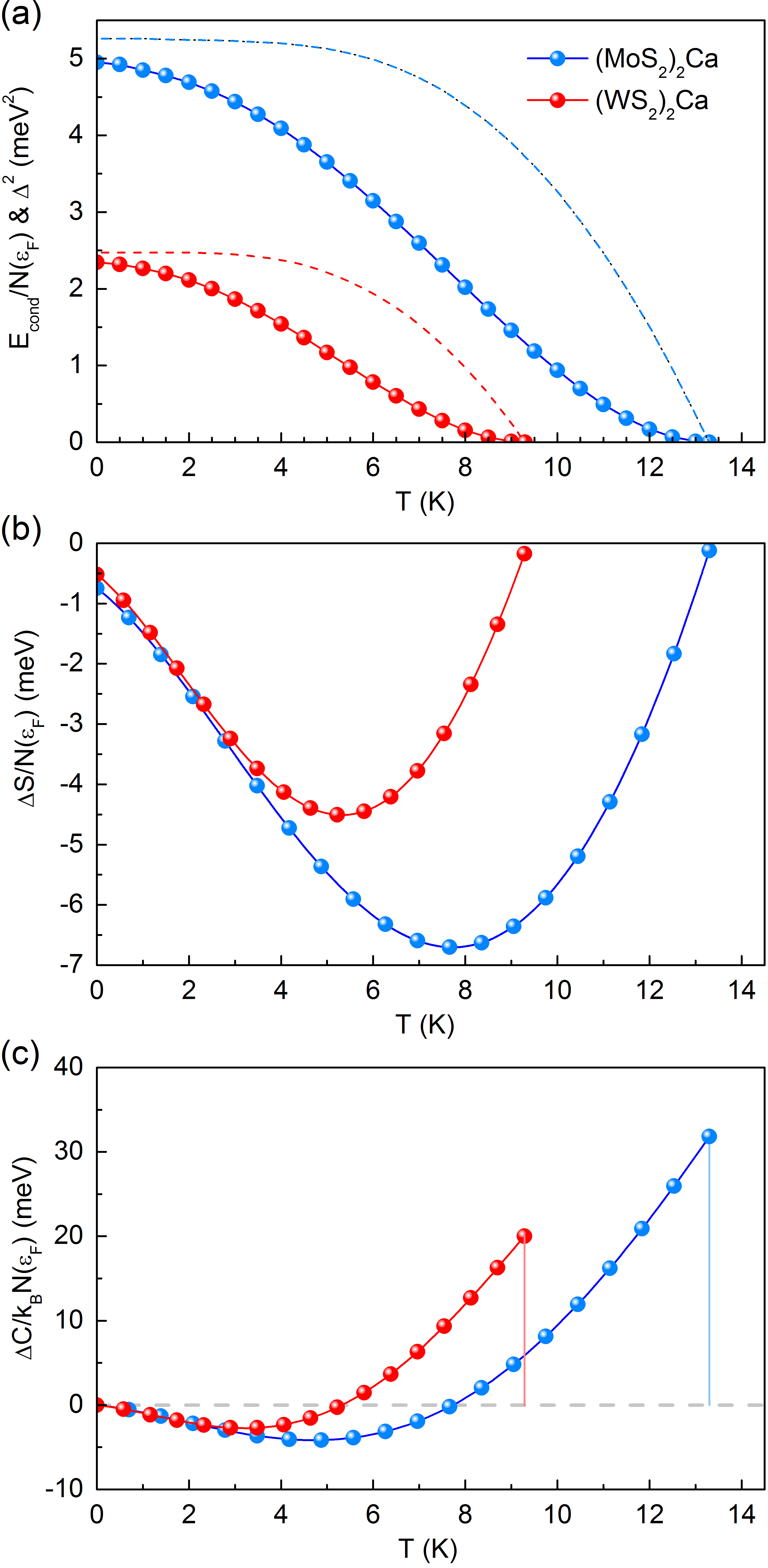} 
\caption{a) Condensation energy (lines with symbols) combined with the energy gap function (dashed lines), b) entropy difference, and c) specific heat difference as a
function of temperature.}         
         \label{f4}
\end{figure}

%%%%%%%%%%%%%%%%%%%%%%%%%%%%%%%%%%%%%%%%%%%%%%%%%%%%%%%%%%%%%%%%%%%%%%%%%%%%%%%%%%%%%%%%%%%%%%%%%%%
From the physical point of view, the condensation energy denotes the energy that stabilizes the superconducting state. 
In \fig{f4}(a), we plot the condensation energy as a function of temperature.
%%%%%%%%%%%%%%%%%%%%%%%%%%%%%%%%%%%%%%%%%%%%%%%%%%%%%%%%%%%%%%%%%%%%%%%%%%%%%%%%%%%%%%%%%%%%%%%%%%%
We should note that at a temperature near zero, an excellent degree of accuracy is achieved with the following relationship $\Delta(0)=\sqrt{E_{\rm cond}(0)}$, where $\Delta(0)$ is an energy gap at the Fermi level. The shapes of functions $\Delta^2(T)$ are plotted in \fig{f4}(a) using a blue dashed line for (MoS$_2$)$_2$Ca and red dashed line for (WS$_2$)$_2$Ca. 
As we can see, with an increase of temperature, the condensation energy decreases and vanishes like an energy gap function at critical temperatures equal to $13.3$ K for (MoS$_2$)$_2$Ca and $9.3$ K for (WS$_2$)$_2$Ca.
Here, we should mention that the recently studied Li-intercalated bilayer MoS$_2$ superconductor is characterized by a slightly lower critical temperature equal to $10.2$ K \cite{HuangMoS}.
In the case of tungsten disulfide, we expect that, even in the absence of experimental or other computational results used for comparison, the calcium is a good intercalant, and we hope that our results can be verified in the future.

Condensation energy is also related to the entropy difference (${\Delta S}$) as well as to the specific heat difference between the superconducting and normal states ($\Delta C$). In particular, ${\Delta S}$ is determined by the first derivative of $E_{\rm cond}$, and $\Delta C$ follows from the second derivative of $E_{\rm cond}$:
\begin{equation}
\label{rS}
\Delta S(T)=\frac{d E_{\rm cond}}{d T} \qquad {\rm and} \qquad \Delta C(T)=T\frac{d^{2}E_{\rm cond}}{dT^{2}}.
\end{equation}

\begin{figure}[!t]         
\includegraphics[width=1\columnwidth]{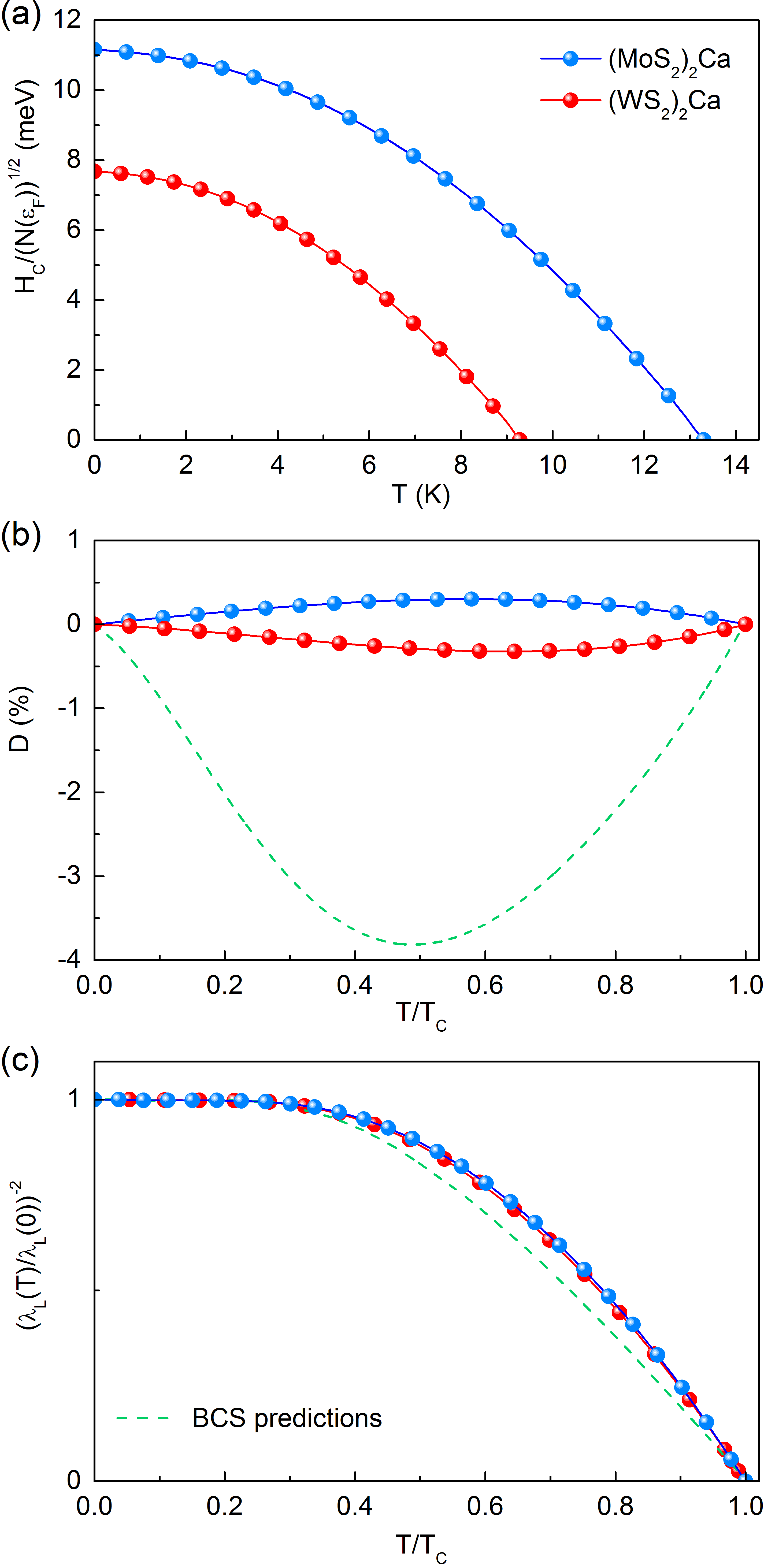} 
\caption{a) Thermodynamic critical field, b) deviation of the thermodynamic critical field,
and c) normalized London penetration depth as a function of temperature.}         
         \label{f5}
\end{figure}

In \fig{f4}(b), we can observe the behavior when an entropy difference appears between the superconducting and normal states. On this basis, we can conclude that the entropy of the superconducting state is lower than that of the normal state because of the ordering of electrons into pairs \cite{Solyom}.
The plot of the specific heat difference between the superconducting and normal states on the temperature is presented in \fig{f4}(c). The characteristic specific heat jump at the critical temperature is marked by the vertical line. 
In the next step, the thermodynamic critical field is calculated ${H_{C}(T)}=\sqrt{8\pi E_{\rm cond}}$ and plotted in \fig{f5}(a).
%%%%%%%%%%%%%%%%%%%%%%%%%%%%%%%%%%%%%%%%%%%%%%%%%%%%%%%%%%%%%%%%%%%%%%%%%%%%%%%%%%%%%%%%%%%%%%%%%%%
%

The estimated values of the energy gap at zero temperature, the specific heat jump at critical temperature, and the thermodynamic critical field at zero temperature allowed us to calculate three fundamental dimensionless ratios: $R_{\Delta}\equiv 2\Delta(0)/k_BT_C$, $R_{C}\equiv{\Delta C\left(T_{C}\right)}/{C^{N}\left(T_{C}\right)}$, and $R_{H}\equiv{T_{C}C^{N}\left(T_{C}\right)}/{H_{C}^{2}\left(0\right)}$, simultaneously. 
Here, the specific heat in the normal state is defined as $C^{N}=\gamma T$, where the Sommerfeld constant has the form: $\gamma\equiv ({2}/{3})\pi^{2}\left(1+\lambda\right)k_{B}^2N(\epsilon_F)$.
Within the framework of the BCS theory \cite{Bardeen1957A, Bardeen1957B}, $R_{\Delta}$, $R_{C}$, and $R_{H}$ ratios adopt universal values equal to $3.53$, $1.43$, and $0.168$, respectively \cite{Carbotte1990A}.
The results obtained for the investigated superconductors are collected in \tab{t1}.
\begin{table}[!h]
\centering
\caption{Dimensionless ratios $R_{\Delta}$, $R_{C}$, $R_{H}$, and $k_BT_C/\omega_{\rm ln}$ of molybdenum and tungsten disulfide bilayers with Ca intercalation. The obtained results are compared with the respective BCS predictions.}
\label{t1}
\begin{ruledtabular}
\begin{tabular}{cccc}
                     & (MoS$_2$)$_2$Ca & (WS$_2$)$_2$Ca  & BCS theory \\\hline
$R_{\Delta}$         &  $4.00$         &   $3.94$        & $3.53$     \\
$R_C$                &  $2.04$         &   $1.89$        & $1.43$     \\
$R_H$                &  $0.145$        &   $0.147$       & $0.168$    \\
$k_BT_C/\omega_{\rm ln}$ & $0.083$ 	       &   $0.070$       & $0$
\end{tabular}
\end{ruledtabular}
\end{table}

We can clearly see that the ratio of energy gap and specific heats as well as the ratio connected with the zero-temperature thermodynamic critical field exceed the predictions of the BCS theory. The discrepancies between our results and the BCS estimates arise from the retardation and strong-coupling effects existing in the studied superconductors. We notice that these effects are well described within the framework of the Eliashberg formalism by the ratio $k_BT_C/\omega_{\rm ln}$, where $\omega_{\rm ln}$ is the logarithmic phonon frequency defined as: $\omega_{{\rm ln}}\equiv \exp\left[\frac{2}{\lambda}
\int^{+\infty}_{0}d\omega\frac{\alpha^{2}F\left(\omega\right)}
{\omega}\ln\left(\omega\right)\right]$. 
The considered ratio equals $0.083$ and $0.070$ for molybdenum and tungsten disulfide bilayers with Ca intercalation, respectively, whereas the mean-field BCS theory is completely omitted (with the weak-coupling BCS limit, we have $k_BT_C/\omega_{ln}\rightarrow 0$).

Derogations from predictions of the BCS theory also can be observed in the shape of the thermodynamic critical field deviation function \cite{Carbotte1990A}:
\begin{equation}
D(T) = \frac{H_C\left(T\right)}{H_C\left(0\right)}-\left[1-\left(\frac{T}{T_C}\right)^2\right]
\label{rD}
\end{equation}
and London penetration depth ($\lambda_{L}$) \cite{Carbotte1990A}:
\begin{eqnarray}
\label{r6}
\frac{1}{e^2v^2_F\rho\left(0\right)\lambda^2_L\left(T\right)}=\frac{4}{3}\frac{\pi}{\beta}\sum_{n=1}^{M}
\frac{\Delta^{2}_{n}}{Z^{S}_{n}\left[ \omega^{2}_{n}+\Delta^{2}_{n}\right]^{3/2}},
\end{eqnarray}
where $e$ and $v^{}_F$ denote the electron charge and Fermi velocity, respectively.
The obtained results compared with the BCS predictions are presented in \fig{f5}(b) and \fig{f5}(c).
%
%%%%%%%%%%%%%%%%%%%%%%%%%%%%%%%%%%%%%%%%%%%%%%%%%%%%%%%%%%%%%%%%%%%%%%%%%%%%%%%%%%%%%%%%%%%%%%%%%%%
\section{Conclusion}
%%%%%%%%%%%%%%%%%%%%%%%%%%%%%%%%%%%%%%%%%%%%%%%%%%%%%%%%%%%%%%%%%%%%%%%%%%%%%%%%%%%%%%%%%%%%%%%%%%%
%
In this study, we performed comprehensive theoretical investigations on the possible existence of a superconducting state in molybdenum and tungsten disulfide bilayers with Ca intercalation.
Combining density functional theory with the Migdal--Eliashberg formalism, we determined that Ca-intercalated bilayer MoS$_2$ and WS$_2$ exhibit a strong-coupling phonon-mediated superconductivity ($\lambda>1$) with $T_C$ of $13.3$ and $9.3$ K, respectively.
In addition, we obtained direct evidence of non-BCS values of dimensionless ratios connected with thermodynamic functions.
In particular, ratios of energy gap, specific heats, and that connected with the thermodynamic critical field significantly exceeded the values predicted by the conventional BCS theory.
We identified these discrepancies when retardation and strong-coupling effects occurred in superconducting Ca-intercalated transition-metal dichalcogenide bilayers.
We expect that (MoS$_2$)$_2$Ca and (WS$_2$)$_2$Ca can be successfully synthesized using chemical or physical methods and applied as nanoscale superconductors.\\

%%%%%%%%%%%%%%%%%%%%%%%%%%%

\noindent{\bf Acknowledgments}

A.P. Durajski acknowledges the financial support under the scholarship START from the Foundation for Polish Science (FNP).
All authors are grateful to the Czestochowa University of Technology---MSK CzestMAN for granting access to the computing infrastructure built in the projects No. POIG.02.03.00-00-028/08 "PLATON - Science Services Platform" and No. POIG.02.03.00-00-110/13 "Deploying high-availability, critical services in Metropolitan Area Networks (MAN-HA)."

%%%%%%%%%%%%%%%%%%%%%%%%%%%
\bibliography{bibliography}

\begin{thebibliography}{53}%
\makeatletter
\providecommand \@ifxundefined [1]{%
 \@ifx{#1\undefined}
}%
\providecommand \@ifnum [1]{%
 \ifnum #1\expandafter \@firstoftwo
 \else \expandafter \@secondoftwo
 \fi
}%
\providecommand \@ifx [1]{%
 \ifx #1\expandafter \@firstoftwo
 \else \expandafter \@secondoftwo
 \fi
}%
\providecommand \natexlab [1]{#1}%
\providecommand \enquote  [1]{``#1''}%
\providecommand \bibnamefont  [1]{#1}%
\providecommand \bibfnamefont [1]{#1}%
\providecommand \citenamefont [1]{#1}%
\providecommand \href@noop [0]{\@secondoftwo}%
\providecommand \href [0]{\begingroup \@sanitize@url \@href}%
\providecommand \@href[1]{\@@startlink{#1}\@@href}%
\providecommand \@@href[1]{\endgroup#1\@@endlink}%
\providecommand \@sanitize@url [0]{\catcode `\\12\catcode `\$12\catcode
  `\&12\catcode `\#12\catcode `\^12\catcode `\_12\catcode `\%12\relax}%
\providecommand \@@startlink[1]{}%
\providecommand \@@endlink[0]{}%
\providecommand \url  [0]{\begingroup\@sanitize@url \@url }%
\providecommand \@url [1]{\endgroup\@href {#1}{\urlprefix }}%
\providecommand \urlprefix  [0]{URL }%
\providecommand \Eprint [0]{\href }%
\providecommand \doibase [0]{http://dx.doi.org/}%
\providecommand \selectlanguage [0]{\@gobble}%
\providecommand \bibinfo  [0]{\@secondoftwo}%
\providecommand \bibfield  [0]{\@secondoftwo}%
\providecommand \translation [1]{[#1]}%
\providecommand \BibitemOpen [0]{}%
\providecommand \bibitemStop [0]{}%
\providecommand \bibitemNoStop [0]{.\EOS\space}%
\providecommand \EOS [0]{\spacefactor3000\relax}%
\providecommand \BibitemShut  [1]{\csname bibitem#1\endcsname}%
\let\auto@bib@innerbib\@empty
%</preamble>
\bibitem [{\citenamefont {Saito}\ \emph
  {et~al.}(2016{\natexlab{a}})\citenamefont {Saito}, \citenamefont {Nojima},\
  and\ \citenamefont {Iwasa}}]{Saito}%
  \BibitemOpen
  \bibfield  {author} {\bibinfo {author} {\bibfnamefont {Yu}~\bibnamefont
  {Saito}}, \bibinfo {author} {\bibfnamefont {Tsutomu}\ \bibnamefont {Nojima}},
  \ and\ \bibinfo {author} {\bibfnamefont {Yoshihiro}\ \bibnamefont {Iwasa}},\
  }\bibfield  {title} {\enquote {\bibinfo {title} {Highly crystalline 2d
  superconductors},}\ }\href@noop {} {\bibfield  {journal} {\bibinfo  {journal}
  {Nat. Rev. Mater.}\ }\textbf {\bibinfo {volume} {2}},\ \bibinfo {pages}
  {16094} (\bibinfo {year} {2016}{\natexlab{a}})}\BibitemShut {NoStop}%
\bibitem [{\citenamefont {Choi}\ \emph {et~al.}(2017)\citenamefont {Choi},
  \citenamefont {Choudhary}, \citenamefont {Han}, \citenamefont {Park},
  \citenamefont {Akinwande},\ and\ \citenamefont {Lee}}]{WonbongChoi}%
  \BibitemOpen
  \bibfield  {author} {\bibinfo {author} {\bibfnamefont {Wonbong}\ \bibnamefont
  {Choi}}, \bibinfo {author} {\bibfnamefont {Nitin}\ \bibnamefont {Choudhary}},
  \bibinfo {author} {\bibfnamefont {Gang~Hee}\ \bibnamefont {Han}}, \bibinfo
  {author} {\bibfnamefont {Juhong}\ \bibnamefont {Park}}, \bibinfo {author}
  {\bibfnamefont {Deji}\ \bibnamefont {Akinwande}}, \ and\ \bibinfo {author}
  {\bibfnamefont {Young~Hee}\ \bibnamefont {Lee}},\ }\bibfield  {title}
  {\enquote {\bibinfo {title} {Recent development of two-dimensional transition
  metal dichalcogenides and their applications},}\ }\href {\doibase
  http://dx.doi.org/10.1016/j.mattod.2016.10.002} {\bibfield  {journal}
  {\bibinfo  {journal} {Mater. Today}\ }\textbf {\bibinfo {volume} {20}},\
  \bibinfo {pages} {116 -- 130} (\bibinfo {year} {2017})}\BibitemShut {NoStop}%
\bibitem [{\citenamefont {Uchihashi}(2017)}]{Uchihashi}%
  \BibitemOpen
  \bibfield  {author} {\bibinfo {author} {\bibfnamefont {Takashi}\ \bibnamefont
  {Uchihashi}},\ }\bibfield  {title} {\enquote {\bibinfo {title}
  {Two-dimensional superconductors with atomic-scale thickness},}\ }\href@noop
  {} {\bibfield  {journal} {\bibinfo  {journal} {Supercond. Sci. Technol.}\
  }\textbf {\bibinfo {volume} {30}},\ \bibinfo {pages} {013002} (\bibinfo
  {year} {2017})}\BibitemShut {NoStop}%
\bibitem [{\citenamefont {Sacepe}\ \emph {et~al.}(2011)\citenamefont {Sacepe},
  \citenamefont {Dubouchet}, \citenamefont {Chapelier}, \citenamefont
  {Sanquer}, \citenamefont {Ovadia}, \citenamefont {Shahar}, \citenamefont
  {Feigelman},\ and\ \citenamefont {Ioffe}}]{Sacepe}%
  \BibitemOpen
  \bibfield  {author} {\bibinfo {author} {\bibfnamefont {Benjamin}\
  \bibnamefont {Sacepe}}, \bibinfo {author} {\bibfnamefont {Thomas}\
  \bibnamefont {Dubouchet}}, \bibinfo {author} {\bibfnamefont {Claude}\
  \bibnamefont {Chapelier}}, \bibinfo {author} {\bibfnamefont {Marc}\
  \bibnamefont {Sanquer}}, \bibinfo {author} {\bibfnamefont {Maoz}\
  \bibnamefont {Ovadia}}, \bibinfo {author} {\bibfnamefont {Dan}\ \bibnamefont
  {Shahar}}, \bibinfo {author} {\bibfnamefont {Mikhail}\ \bibnamefont
  {Feigelman}}, \ and\ \bibinfo {author} {\bibfnamefont {Lev}\ \bibnamefont
  {Ioffe}},\ }\bibfield  {title} {\enquote {\bibinfo {title} {Localization of
  preformed cooper pairs in disordered superconductors},}\ }\href {\doibase
  10.1038/nphys1892} {\bibfield  {journal} {\bibinfo  {journal} {Nat. Phys.}\
  }\textbf {\bibinfo {volume} {7}},\ \bibinfo {pages} {239–244} (\bibinfo
  {year} {2011})}\BibitemShut {NoStop}%
\bibitem [{\citenamefont {Guo}\ \emph {et~al.}(2004)\citenamefont {Guo},
  \citenamefont {Zhang}, \citenamefont {Bao}, \citenamefont {Han},
  \citenamefont {Tang}, \citenamefont {Zhang}, \citenamefont {Zhu},
  \citenamefont {Wang}, \citenamefont {Niu}, \citenamefont {Qiu}, \citenamefont
  {Jia}, \citenamefont {Zhao},\ and\ \citenamefont {Xue}}]{GuoYang}%
  \BibitemOpen
  \bibfield  {author} {\bibinfo {author} {\bibfnamefont {Yang}\ \bibnamefont
  {Guo}}, \bibinfo {author} {\bibfnamefont {Yan-Feng}\ \bibnamefont {Zhang}},
  \bibinfo {author} {\bibfnamefont {Xin-Yu}\ \bibnamefont {Bao}}, \bibinfo
  {author} {\bibfnamefont {Tie-Zhu}\ \bibnamefont {Han}}, \bibinfo {author}
  {\bibfnamefont {Zhe}\ \bibnamefont {Tang}}, \bibinfo {author} {\bibfnamefont
  {Li-Xin}\ \bibnamefont {Zhang}}, \bibinfo {author} {\bibfnamefont
  {Wen-Guang}\ \bibnamefont {Zhu}}, \bibinfo {author} {\bibfnamefont {E.~G.}\
  \bibnamefont {Wang}}, \bibinfo {author} {\bibfnamefont {Qian}\ \bibnamefont
  {Niu}}, \bibinfo {author} {\bibfnamefont {Z.~Q.}\ \bibnamefont {Qiu}},
  \bibinfo {author} {\bibfnamefont {Jin-Feng}\ \bibnamefont {Jia}}, \bibinfo
  {author} {\bibfnamefont {Zhong-Xian}\ \bibnamefont {Zhao}}, \ and\ \bibinfo
  {author} {\bibfnamefont {Qi-Kun}\ \bibnamefont {Xue}},\ }\bibfield  {title}
  {\enquote {\bibinfo {title} {Superconductivity modulated by quantum size
  effects},}\ }\href@noop {} {\bibfield  {journal} {\bibinfo  {journal}
  {Science}\ }\textbf {\bibinfo {volume} {306}},\ \bibinfo {pages} {1915--1917}
  (\bibinfo {year} {2004})}\BibitemShut {NoStop}%
\bibitem [{\citenamefont {Durajski}(2015{\natexlab{a}})}]{DurajskiPb}%
  \BibitemOpen
  \bibfield  {author} {\bibinfo {author} {\bibfnamefont {A~P}\ \bibnamefont
  {Durajski}},\ }\bibfield  {title} {\enquote {\bibinfo {title} {Effect of
  layer thickness on the superconducting properties in ultrathin pb films},}\
  }\href@noop {} {\bibfield  {journal} {\bibinfo  {journal} {Supercond. Sci.
  Technol.}\ }\textbf {\bibinfo {volume} {28}},\ \bibinfo {pages} {095011}
  (\bibinfo {year} {2015}{\natexlab{a}})}\BibitemShut {NoStop}%
\bibitem [{\citenamefont {Talantsev}\ \emph {et~al.}(2017)\citenamefont
  {Talantsev}, \citenamefont {Crump}, \citenamefont {Island}, \citenamefont
  {Xing}, \citenamefont {Sun}, \citenamefont {Wang},\ and\ \citenamefont
  {Tallon}}]{Talantsev}%
  \BibitemOpen
  \bibfield  {author} {\bibinfo {author} {\bibfnamefont {E~F}\ \bibnamefont
  {Talantsev}}, \bibinfo {author} {\bibfnamefont {W~P}\ \bibnamefont {Crump}},
  \bibinfo {author} {\bibfnamefont {J~O}\ \bibnamefont {Island}}, \bibinfo
  {author} {\bibfnamefont {Ying}\ \bibnamefont {Xing}}, \bibinfo {author}
  {\bibfnamefont {Yi}~\bibnamefont {Sun}}, \bibinfo {author} {\bibfnamefont
  {Jian}\ \bibnamefont {Wang}}, \ and\ \bibinfo {author} {\bibfnamefont {J~L}\
  \bibnamefont {Tallon}},\ }\bibfield  {title} {\enquote {\bibinfo {title} {On
  the origin of critical temperature enhancement in atomically thin
  superconductors},}\ }\href@noop {} {\bibfield  {journal} {\bibinfo  {journal}
  {2D Mater.}\ }\textbf {\bibinfo {volume} {4}},\ \bibinfo {pages} {025072}
  (\bibinfo {year} {2017})}\BibitemShut {NoStop}%
\bibitem [{\citenamefont {Goldman}\ and\ \citenamefont
  {Markovic}(1998)}]{GoldmanMarkovic}%
  \BibitemOpen
  \bibfield  {author} {\bibinfo {author} {\bibfnamefont {Allen~M.}\
  \bibnamefont {Goldman}}\ and\ \bibinfo {author} {\bibfnamefont {Nina}\
  \bibnamefont {Markovic}},\ }\bibfield  {title} {\enquote {\bibinfo {title}
  {Superconductor-insulator transitions in the two-dimensional limit},}\ }\href
  {\doibase 10.1063/1.882069} {\bibfield  {journal} {\bibinfo  {journal} {Phys.
  Today}\ }\textbf {\bibinfo {volume} {51}},\ \bibinfo {pages} {39} (\bibinfo
  {year} {1998})}\BibitemShut {NoStop}%
\bibitem [{\citenamefont {Berezinskii}(1971)}]{Berezinskii1}%
  \BibitemOpen
  \bibfield  {author} {\bibinfo {author} {\bibfnamefont { V. L.}\ \bibnamefont
  {Berezinskii}},\ }\bibfield  {title} {\enquote {\bibinfo {title} {Destruction
  of long-range order in one-dimensional and two-dimensional systems having a
  continuous symmetry group. {I.} classical systems},}\ }\href@noop {}
  {\bibfield  {journal} {\bibinfo  {journal} {Sov. Phys. JETP}\ }\textbf
  {\bibinfo {volume} {32}},\ \bibinfo {pages} {493–500} (\bibinfo {year}
  {1971})}\BibitemShut {NoStop}%
\bibitem [{\citenamefont {Berezinskii}(1972)}]{Berezinskii2}%
  \BibitemOpen
  \bibfield  {author} {\bibinfo {author} {\bibfnamefont { V. L.}\ \bibnamefont
  {Berezinskii}},\ }\bibfield  {title} {\enquote {\bibinfo {title} {Destruction
  of long-range order in one-dimensional and two-dimensional systems possessing
  a continuous symmetry group. {II.} quantum systems},}\ }\href@noop {}
  {\bibfield  {journal} {\bibinfo  {journal} {Sov. Phys. JETP}\ }\textbf
  {\bibinfo {volume} {34}},\ \bibinfo {pages} {610–616} (\bibinfo {year}
  {1972})}\BibitemShut {NoStop}%
\bibitem [{\citenamefont {Kosterlitz}\ and\ \citenamefont
  {Thouless}(1972)}]{Berezinskii3}%
  \BibitemOpen
  \bibfield  {author} {\bibinfo {author} {\bibfnamefont { J. M.}\ \bibnamefont
  {Kosterlitz}}\ and\ \bibinfo {author} {\bibfnamefont { D. J.}\ \bibnamefont
  {Thouless}},\ }\bibfield  {title} {\enquote {\bibinfo {title} {Long range
  order and metastability in two dimensional solids and superfluids},}\
  }\href@noop {} {\bibfield  {journal} {\bibinfo  {journal} {J. Phys. C Solid
  State Phys.}\ }\textbf {\bibinfo {volume} {5}},\ \bibinfo {pages} {124–126}
  (\bibinfo {year} {1972})}\BibitemShut {NoStop}%
\bibitem [{\citenamefont {Zhang}\ \emph {et~al.}(2017)\citenamefont {Zhang},
  \citenamefont {Wu}, \citenamefont {Yang},\ and\ \citenamefont
  {Zhang}}]{Zhang2017}%
  \BibitemOpen
  \bibfield  {author} {\bibinfo {author} {\bibfnamefont {Tingting}\
  \bibnamefont {Zhang}}, \bibinfo {author} {\bibfnamefont {Shuang}\
  \bibnamefont {Wu}}, \bibinfo {author} {\bibfnamefont {Rong}\ \bibnamefont
  {Yang}}, \ and\ \bibinfo {author} {\bibfnamefont {Guangyu}\ \bibnamefont
  {Zhang}},\ }\bibfield  {title} {\enquote {\bibinfo {title} {Graphene:
  Nanostructure engineering and applications},}\ }\href {\doibase
  10.1007/s11467-017-0648-z} {\bibfield  {journal} {\bibinfo  {journal} {Front.
  Phys.}\ }\textbf {\bibinfo {volume} {12}},\ \bibinfo {pages} {127206}
  (\bibinfo {year} {2017})}\BibitemShut {NoStop}%
\bibitem [{\citenamefont {Wang}\ \emph {et~al.}(2012)\citenamefont {Wang},
  \citenamefont {Kalantar-Zadeh}, \citenamefont {Kis}, \citenamefont
  {Coleman},\ and\ \citenamefont {Strano}}]{HuaWang}%
  \BibitemOpen
  \bibfield  {author} {\bibinfo {author} {\bibfnamefont {Qing~Hua}\
  \bibnamefont {Wang}}, \bibinfo {author} {\bibfnamefont {Kourosh}\
  \bibnamefont {Kalantar-Zadeh}}, \bibinfo {author} {\bibfnamefont {Andras}\
  \bibnamefont {Kis}}, \bibinfo {author} {\bibfnamefont {Jonathan~N.}\
  \bibnamefont {Coleman}}, \ and\ \bibinfo {author} {\bibfnamefont
  {Michael~S.}\ \bibnamefont {Strano}},\ }\bibfield  {title} {\enquote
  {\bibinfo {title} {Electronics and optoelectronics of two-dimensional
  transition metal dichalcogenides},}\ }\href@noop {} {\bibfield  {journal}
  {\bibinfo  {journal} {Nat. Nanotechnol.}\ }\textbf {\bibinfo {volume} {7}},\
  \bibinfo {pages} {699–712} (\bibinfo {year} {2012})}\BibitemShut {NoStop}%
\bibitem [{\citenamefont {Duan}\ \emph {et~al.}(2015)\citenamefont {Duan},
  \citenamefont {Wang}, \citenamefont {Pan}, \citenamefont {Yu},\ and\
  \citenamefont {Duan}}]{DuanXidong}%
  \BibitemOpen
  \bibfield  {author} {\bibinfo {author} {\bibfnamefont {Xidong}\ \bibnamefont
  {Duan}}, \bibinfo {author} {\bibfnamefont {Chen}\ \bibnamefont {Wang}},
  \bibinfo {author} {\bibfnamefont {Anlian}\ \bibnamefont {Pan}}, \bibinfo
  {author} {\bibfnamefont {Ruqin}\ \bibnamefont {Yu}}, \ and\ \bibinfo {author}
  {\bibfnamefont {Xiangfeng}\ \bibnamefont {Duan}},\ }\bibfield  {title}
  {\enquote {\bibinfo {title} {Two-dimensional transition metal dichalcogenides
  as atomically thin semiconductors: opportunities and challenges},}\
  }\href@noop {} {\bibfield  {journal} {\bibinfo  {journal} {Chem. Soc. Rev.}\
  }\textbf {\bibinfo {volume} {44}},\ \bibinfo {pages} {8859--8876} (\bibinfo
  {year} {2015})}\BibitemShut {NoStop}%
\bibitem [{\citenamefont {Guo}\ and\ \citenamefont
  {Robertson}(2016)}]{YuzhengGuo}%
  \BibitemOpen
  \bibfield  {author} {\bibinfo {author} {\bibfnamefont {Yuzheng}\ \bibnamefont
  {Guo}}\ and\ \bibinfo {author} {\bibfnamefont {John}\ \bibnamefont
  {Robertson}},\ }\bibfield  {title} {\enquote {\bibinfo {title} {Band
  engineering in transition metal dichalcogenides: Stacked versus lateral
  heterostructures},}\ }\href@noop {} {\bibfield  {journal} {\bibinfo
  {journal} {Appl. Phys. Lett.}\ }\textbf {\bibinfo {volume} {108}},\ \bibinfo
  {pages} {233104} (\bibinfo {year} {2016})}\BibitemShut {NoStop}%
\bibitem [{\citenamefont {Szcz{\c{e}}{\'s}niak}\ \emph
  {et~al.}(2016)\citenamefont {Szcz{\c{e}}{\'s}niak}, \citenamefont {Ennaoui},\
  and\ \citenamefont {Ahzi}}]{DominMoS}%
  \BibitemOpen
  \bibfield  {author} {\bibinfo {author} {\bibfnamefont {Dominik}\ \bibnamefont
  {Szcz{\c{e}}{\'s}niak}}, \bibinfo {author} {\bibfnamefont {Ahmed}\
  \bibnamefont {Ennaoui}}, \ and\ \bibinfo {author} {\bibfnamefont {Said}\
  \bibnamefont {Ahzi}},\ }\bibfield  {title} {\enquote {\bibinfo {title}
  {Complex band structures of transition metal dichalcogenide monolayers with
  spin-orbit coupling effects},}\ }\href@noop {} {\bibfield  {journal}
  {\bibinfo  {journal} {J. Phys.: Condens. Matter}\ }\textbf {\bibinfo {volume}
  {28}},\ \bibinfo {pages} {355301} (\bibinfo {year} {2016})}\BibitemShut
  {NoStop}%
\bibitem [{\citenamefont {Kandemir}\ \emph {et~al.}(2016)\citenamefont
  {Kandemir}, \citenamefont {Yapicioglu}, \citenamefont {Kinaci}, \citenamefont
  {Çağın},\ and\ \citenamefont {Sevik}}]{AliKandemir}%
  \BibitemOpen
  \bibfield  {author} {\bibinfo {author} {\bibfnamefont {Ali}\ \bibnamefont
  {Kandemir}}, \bibinfo {author} {\bibfnamefont {Haluk}\ \bibnamefont
  {Yapicioglu}}, \bibinfo {author} {\bibfnamefont {Alper}\ \bibnamefont
  {Kinaci}}, \bibinfo {author} {\bibfnamefont {Tahir}\ \bibnamefont
  {Çağın}}, \ and\ \bibinfo {author} {\bibfnamefont {Cem}\ \bibnamefont
  {Sevik}},\ }\bibfield  {title} {\enquote {\bibinfo {title} {Thermal transport
  properties of {MoS$_2$} and {MoSe$_2$} monolayers},}\ }\href@noop {}
  {\bibfield  {journal} {\bibinfo  {journal} {Nanotechnology}\ }\textbf
  {\bibinfo {volume} {27}},\ \bibinfo {pages} {055703} (\bibinfo {year}
  {2016})}\BibitemShut {NoStop}%
\bibitem [{\citenamefont {Zhao}\ \emph {et~al.}(2017)\citenamefont {Zhao},
  \citenamefont {Zheng}, \citenamefont {Guo}, \citenamefont {Jiang},
  \citenamefont {Cao},\ and\ \citenamefont {Wan}}]{PujuZhao}%
  \BibitemOpen
  \bibfield  {author} {\bibinfo {author} {\bibfnamefont {Puju}\ \bibnamefont
  {Zhao}}, \bibinfo {author} {\bibfnamefont {Jiming}\ \bibnamefont {Zheng}},
  \bibinfo {author} {\bibfnamefont {Ping}\ \bibnamefont {Guo}}, \bibinfo
  {author} {\bibfnamefont {Zhenyi}\ \bibnamefont {Jiang}}, \bibinfo {author}
  {\bibfnamefont {Like}\ \bibnamefont {Cao}}, \ and\ \bibinfo {author}
  {\bibfnamefont {Yun}\ \bibnamefont {Wan}},\ }\bibfield  {title} {\enquote
  {\bibinfo {title} {Electronic and magnetic properties of re-doped
  single-layer {MoS$_2$}: A {DFT} study},}\ }\href@noop {} {\bibfield
  {journal} {\bibinfo  {journal} {Comput. Mater. Sci.}\ }\textbf {\bibinfo
  {volume} {128}},\ \bibinfo {pages} {287 -- 293} (\bibinfo {year}
  {2017})}\BibitemShut {NoStop}%
\bibitem [{\citenamefont {Kam}\ and\ \citenamefont
  {Parkinson}(1982)}]{KamMoS2}%
  \BibitemOpen
  \bibfield  {author} {\bibinfo {author} {\bibfnamefont {K.~K.}\ \bibnamefont
  {Kam}}\ and\ \bibinfo {author} {\bibfnamefont {B.~A.}\ \bibnamefont
  {Parkinson}},\ }\bibfield  {title} {\enquote {\bibinfo {title} {Detailed
  photocurrent spectroscopy of the semiconducting group vib transition metal
  dichalcogenides},}\ }\href@noop {} {\bibfield  {journal} {\bibinfo  {journal}
  {J. Phys. Chem.}\ }\textbf {\bibinfo {volume} {86}},\ \bibinfo {pages}
  {463--467} (\bibinfo {year} {1982})}\BibitemShut {NoStop}%
\bibitem [{\citenamefont {Mak}\ \emph {et~al.}(2010)\citenamefont {Mak},
  \citenamefont {Lee}, \citenamefont {Hone}, \citenamefont {Shan},\ and\
  \citenamefont {Heinz}}]{MakMoS2}%
  \BibitemOpen
  \bibfield  {author} {\bibinfo {author} {\bibfnamefont {Kin~Fai}\ \bibnamefont
  {Mak}}, \bibinfo {author} {\bibfnamefont {Changgu}\ \bibnamefont {Lee}},
  \bibinfo {author} {\bibfnamefont {James}\ \bibnamefont {Hone}}, \bibinfo
  {author} {\bibfnamefont {Jie}\ \bibnamefont {Shan}}, \ and\ \bibinfo {author}
  {\bibfnamefont {Tony~F.}\ \bibnamefont {Heinz}},\ }\bibfield  {title}
  {\enquote {\bibinfo {title} {Atomically thin {MoS$_2$}: A new direct-gap
  semiconductor},}\ }\href@noop {} {\bibfield  {journal} {\bibinfo  {journal}
  {Phys. Rev. Lett.}\ }\textbf {\bibinfo {volume} {105}},\ \bibinfo {pages}
  {136805} (\bibinfo {year} {2010})}\BibitemShut {NoStop}%
\bibitem [{\citenamefont {Luo}\ \emph {et~al.}(2017)\citenamefont {Luo},
  \citenamefont {Zhang}, \citenamefont {Li}, \citenamefont {Song},
  \citenamefont {Deng}, \citenamefont {Cao}, \citenamefont {Xiao},\ and\
  \citenamefont {Guo}}]{Luo2017}%
  \BibitemOpen
  \bibfield  {author} {\bibinfo {author} {\bibfnamefont {Gang}\ \bibnamefont
  {Luo}}, \bibinfo {author} {\bibfnamefont {Zhuo-Zhi}\ \bibnamefont {Zhang}},
  \bibinfo {author} {\bibfnamefont {Hai-Ou}\ \bibnamefont {Li}}, \bibinfo
  {author} {\bibfnamefont {Xiang-Xiang}\ \bibnamefont {Song}}, \bibinfo
  {author} {\bibfnamefont {Guang-Wei}\ \bibnamefont {Deng}}, \bibinfo {author}
  {\bibfnamefont {Gang}\ \bibnamefont {Cao}}, \bibinfo {author} {\bibfnamefont
  {Ming}\ \bibnamefont {Xiao}}, \ and\ \bibinfo {author} {\bibfnamefont
  {Guo-Ping}\ \bibnamefont {Guo}},\ }\bibfield  {title} {\enquote {\bibinfo
  {title} {Quantum dot behavior in transition metal dichalcogenides
  nanostructures},}\ }\href {\doibase 10.1007/s11467-017-0652-3} {\bibfield
  {journal} {\bibinfo  {journal} {Front. Phys.}\ }\textbf {\bibinfo {volume}
  {12}},\ \bibinfo {pages} {128502} (\bibinfo {year} {2017})}\BibitemShut
  {NoStop}%
\bibitem [{\citenamefont {Novoselov}\ \emph {et~al.}(2004)\citenamefont
  {Novoselov}, \citenamefont {Geim}, \citenamefont {Morozov}, \citenamefont
  {Jiang}, \citenamefont {Zhang}, \citenamefont {Dubonos}, \citenamefont
  {Grigorieva},\ and\ \citenamefont {Firsov}}]{novoselov}%
  \BibitemOpen
  \bibfield  {author} {\bibinfo {author} {\bibfnamefont {K.~S.}\ \bibnamefont
  {Novoselov}}, \bibinfo {author} {\bibfnamefont {A.~K.}\ \bibnamefont {Geim}},
  \bibinfo {author} {\bibfnamefont {S.~V.}\ \bibnamefont {Morozov}}, \bibinfo
  {author} {\bibfnamefont {D.}~\bibnamefont {Jiang}}, \bibinfo {author}
  {\bibfnamefont {Y.}~\bibnamefont {Zhang}}, \bibinfo {author} {\bibfnamefont
  {S.~V.}\ \bibnamefont {Dubonos}}, \bibinfo {author} {\bibfnamefont {I.~V.}\
  \bibnamefont {Grigorieva}}, \ and\ \bibinfo {author} {\bibfnamefont {A.~A.}\
  \bibnamefont {Firsov}},\ }\bibfield  {title} {\enquote {\bibinfo {title}
  {Electric field effect in atomically thin carbon films},}\ }\href@noop {}
  {\bibfield  {journal} {\bibinfo  {journal} {Science}\ }\textbf {\bibinfo
  {volume} {306}},\ \bibinfo {pages} {666--669} (\bibinfo {year}
  {2004})}\BibitemShut {NoStop}%
\bibitem [{\citenamefont {Jiang}(2015)}]{JiangMoS2Graphene}%
  \BibitemOpen
  \bibfield  {author} {\bibinfo {author} {\bibfnamefont {Jin-Wu}\ \bibnamefont
  {Jiang}},\ }\bibfield  {title} {\enquote {\bibinfo {title} {Graphene versus
  {MoS$_2$}: A short review},}\ }\href@noop {} {\bibfield  {journal} {\bibinfo
  {journal} {Front. Phys.}\ }\textbf {\bibinfo {volume} {10}},\ \bibinfo
  {pages} {287--302} (\bibinfo {year} {2015})}\BibitemShut {NoStop}%
\bibitem [{\citenamefont {Durajski}(2015{\natexlab{b}})}]{DurajskiGraphene}%
  \BibitemOpen
  \bibfield  {author} {\bibinfo {author} {\bibfnamefont {A~P}\ \bibnamefont
  {Durajski}},\ }\bibfield  {title} {\enquote {\bibinfo {title} {Influence of
  hole doping on the superconducting state in graphane},}\ }\href@noop {}
  {\bibfield  {journal} {\bibinfo  {journal} {Supercond. Sci. Technol.}\
  }\textbf {\bibinfo {volume} {28}},\ \bibinfo {pages} {035002} (\bibinfo
  {year} {2015}{\natexlab{b}})}\BibitemShut {NoStop}%
\bibitem [{\citenamefont {Lin}\ \emph {et~al.}(2016)\citenamefont {Lin},
  \citenamefont {Li}, \citenamefont {Dong}, \citenamefont {Wang}, \citenamefont
  {Chen},\ and\ \citenamefont {Zhang}}]{XinyueLin}%
  \BibitemOpen
  \bibfield  {author} {\bibinfo {author} {\bibfnamefont {Xinyue}\ \bibnamefont
  {Lin}}, \bibinfo {author} {\bibfnamefont {Wentong}\ \bibnamefont {Li}},
  \bibinfo {author} {\bibfnamefont {Yingying}\ \bibnamefont {Dong}}, \bibinfo
  {author} {\bibfnamefont {Chen}\ \bibnamefont {Wang}}, \bibinfo {author}
  {\bibfnamefont {Qi}~\bibnamefont {Chen}}, \ and\ \bibinfo {author}
  {\bibfnamefont {Hui}\ \bibnamefont {Zhang}},\ }\bibfield  {title} {\enquote
  {\bibinfo {title} {Two-dimensional metallic {MoS$_2$}: A dft study},}\
  }\href@noop {} {\bibfield  {journal} {\bibinfo  {journal} {Comput. Mater.
  Sci.}\ }\textbf {\bibinfo {volume} {124}},\ \bibinfo {pages} {49 -- 53}
  (\bibinfo {year} {2016})}\BibitemShut {NoStop}%
\bibitem [{\citenamefont {Pesic}\ \emph {et~al.}(2014)\citenamefont {Pesic},
  \citenamefont {Gajic}, \citenamefont {Hingerl},\ and\ \citenamefont
  {Belic}}]{JelenaPesic}%
  \BibitemOpen
  \bibfield  {author} {\bibinfo {author} {\bibfnamefont {Jelena}\ \bibnamefont
  {Pesic}}, \bibinfo {author} {\bibfnamefont {R}~\bibnamefont {Gajic}},
  \bibinfo {author} {\bibfnamefont {Kurt}\ \bibnamefont {Hingerl}}, \ and\
  \bibinfo {author} {\bibfnamefont {Milivoj}\ \bibnamefont {Belic}},\
  }\bibfield  {title} {\enquote {\bibinfo {title} {Strain-enhanced
  superconductivity in {Li}-doped graphene},}\ }\href@noop {} {\bibfield
  {journal} {\bibinfo  {journal} {Europhys. Lett.}\ }\textbf {\bibinfo {volume}
  {108}},\ \bibinfo {pages} {67005} (\bibinfo {year} {2014})}\BibitemShut
  {NoStop}%
\bibitem [{\citenamefont {Zhang}\ \emph {et~al.}(2016)\citenamefont {Zhang},
  \citenamefont {Gao},\ and\ \citenamefont {Dong}}]{ZhangMoS}%
  \BibitemOpen
  \bibfield  {author} {\bibinfo {author} {\bibfnamefont {Jun-Jie}\ \bibnamefont
  {Zhang}}, \bibinfo {author} {\bibfnamefont {Bin}\ \bibnamefont {Gao}}, \ and\
  \bibinfo {author} {\bibfnamefont {Shuai}\ \bibnamefont {Dong}},\ }\bibfield
  {title} {\enquote {\bibinfo {title} {Strain-enhanced superconductivity of
  $\mathrm{Mo}{X}_{2}(x=\text{S} \text{ or Se})$ bilayers with na
  intercalation},}\ }\href@noop {} {\bibfield  {journal} {\bibinfo  {journal}
  {Phys. Rev. B}\ }\textbf {\bibinfo {volume} {93}},\ \bibinfo {pages} {155430}
  (\bibinfo {year} {2016})}\BibitemShut {NoStop}%
\bibitem [{\citenamefont {Huang}\ \emph {et~al.}(2016)\citenamefont {Huang},
  \citenamefont {Xing},\ and\ \citenamefont {Xing}}]{HuangMoS}%
  \BibitemOpen
  \bibfield  {author} {\bibinfo {author} {\bibfnamefont {G.~Q.}\ \bibnamefont
  {Huang}}, \bibinfo {author} {\bibfnamefont {Z.~W.}\ \bibnamefont {Xing}}, \
  and\ \bibinfo {author} {\bibfnamefont {D.~Y.}\ \bibnamefont {Xing}},\
  }\bibfield  {title} {\enquote {\bibinfo {title} {Dynamical stability and
  superconductivity of {Li}-intercalated bilayer {MoS$_2$}: A first-principles
  prediction},}\ }\href@noop {} {\bibfield  {journal} {\bibinfo  {journal}
  {Phys. Rev. B}\ }\textbf {\bibinfo {volume} {93}},\ \bibinfo {pages} {104511}
  (\bibinfo {year} {2016})}\BibitemShut {NoStop}%
\bibitem [{\citenamefont {He}\ \emph {et~al.}(2016)\citenamefont {He},
  \citenamefont {Li}, \citenamefont {Zhu}, \citenamefont {Dai}, \citenamefont
  {Yang}, \citenamefont {Yang}, \citenamefont {Zhang}, \citenamefont {Li},
  \citenamefont {Schwingenschlogl},\ and\ \citenamefont {Zhang}}]{XinHe}%
  \BibitemOpen
  \bibfield  {author} {\bibinfo {author} {\bibfnamefont {Xin}\ \bibnamefont
  {He}}, \bibinfo {author} {\bibfnamefont {Hai}\ \bibnamefont {Li}}, \bibinfo
  {author} {\bibfnamefont {Zhiyong}\ \bibnamefont {Zhu}}, \bibinfo {author}
  {\bibfnamefont {Zhenyu}\ \bibnamefont {Dai}}, \bibinfo {author}
  {\bibfnamefont {Yang}\ \bibnamefont {Yang}}, \bibinfo {author} {\bibfnamefont
  {Peng}\ \bibnamefont {Yang}}, \bibinfo {author} {\bibfnamefont {Qiang}\
  \bibnamefont {Zhang}}, \bibinfo {author} {\bibfnamefont {Peng}\ \bibnamefont
  {Li}}, \bibinfo {author} {\bibfnamefont {Udo}\ \bibnamefont
  {Schwingenschlogl}}, \ and\ \bibinfo {author} {\bibfnamefont {Xixiang}\
  \bibnamefont {Zhang}},\ }\bibfield  {title} {\enquote {\bibinfo {title}
  {Strain engineering in monolayer {WS$_2$}, {MoS$_2$}, and the
  {WS$_2$}/{MoS$_2$} heterostructure},}\ }\href@noop {} {\bibfield  {journal}
  {\bibinfo  {journal} {Appl. Phys. Lett.}\ }\textbf {\bibinfo {volume}
  {109}},\ \bibinfo {pages} {173105} (\bibinfo {year} {2016})}\BibitemShut
  {NoStop}%
\bibitem [{\citenamefont {Ye}\ \emph {et~al.}(2012)\citenamefont {Ye},
  \citenamefont {Zhang}, \citenamefont {Akashi}, \citenamefont {Bahramy},
  \citenamefont {Arita},\ and\ \citenamefont {Iwasa}}]{Ye1193}%
  \BibitemOpen
  \bibfield  {author} {\bibinfo {author} {\bibfnamefont {J.~T.}\ \bibnamefont
  {Ye}}, \bibinfo {author} {\bibfnamefont {Y.~J.}\ \bibnamefont {Zhang}},
  \bibinfo {author} {\bibfnamefont {R.}~\bibnamefont {Akashi}}, \bibinfo
  {author} {\bibfnamefont {M.~S.}\ \bibnamefont {Bahramy}}, \bibinfo {author}
  {\bibfnamefont {R.}~\bibnamefont {Arita}}, \ and\ \bibinfo {author}
  {\bibfnamefont {Y.}~\bibnamefont {Iwasa}},\ }\bibfield  {title} {\enquote
  {\bibinfo {title} {Superconducting dome in a gate-tuned band insulator},}\
  }\href@noop {} {\bibfield  {journal} {\bibinfo  {journal} {Science}\ }\textbf
  {\bibinfo {volume} {338}},\ \bibinfo {pages} {1193--1196} (\bibinfo {year}
  {2012})}\BibitemShut {NoStop}%
\bibitem [{\citenamefont {Nayak}\ \emph {et~al.}(2014)\citenamefont {Nayak},
  \citenamefont {Bhattacharyya}, \citenamefont {Zhu}, \citenamefont {Liu},
  \citenamefont {Wu}, \citenamefont {Pandey}, \citenamefont {Jin},
  \citenamefont {Singh}, \citenamefont {Akinwande},\ and\ \citenamefont
  {Lin}}]{Nayak}%
  \BibitemOpen
  \bibfield  {author} {\bibinfo {author} {\bibfnamefont {Avinash~P.}\
  \bibnamefont {Nayak}}, \bibinfo {author} {\bibfnamefont {Swastibrata}\
  \bibnamefont {Bhattacharyya}}, \bibinfo {author} {\bibfnamefont {Jie}\
  \bibnamefont {Zhu}}, \bibinfo {author} {\bibfnamefont {Jin}\ \bibnamefont
  {Liu}}, \bibinfo {author} {\bibfnamefont {Xiang}\ \bibnamefont {Wu}},
  \bibinfo {author} {\bibfnamefont {Tribhuwan}\ \bibnamefont {Pandey}},
  \bibinfo {author} {\bibfnamefont {Changqing}\ \bibnamefont {Jin}}, \bibinfo
  {author} {\bibfnamefont {Abhishek~K.}\ \bibnamefont {Singh}}, \bibinfo
  {author} {\bibfnamefont {Deji}\ \bibnamefont {Akinwande}}, \ and\ \bibinfo
  {author} {\bibfnamefont {Jung-Fu}\ \bibnamefont {Lin}},\ }\bibfield  {title}
  {\enquote {\bibinfo {title} {Pressure-induced semiconducting to metallic
  transition in multilayered molybdenum disulphide},}\ }\href@noop {}
  {\bibfield  {journal} {\bibinfo  {journal} {Nature Commun.}\ }\textbf
  {\bibinfo {volume} {5}},\ \bibinfo {pages} {3731} (\bibinfo {year}
  {2014})}\BibitemShut {NoStop}%
\bibitem [{\citenamefont {Chi}\ \emph {et~al.}(2015)\citenamefont {Chi},
  \citenamefont {Yen}, \citenamefont {Peng}, \citenamefont {Zhu}, \citenamefont
  {Zhang}, \citenamefont {Chen}, \citenamefont {Yang}, \citenamefont {Liu},
  \citenamefont {Ma}, \citenamefont {Zhao}, \citenamefont {Kagayama},\ and\
  \citenamefont {Iwasa}}]{ZhenhuaChi}%
  \BibitemOpen
  \bibfield  {author} {\bibinfo {author} {\bibfnamefont {Zhenhua}\ \bibnamefont
  {Chi}}, \bibinfo {author} {\bibfnamefont {Feihsiang}\ \bibnamefont {Yen}},
  \bibinfo {author} {\bibfnamefont {Feng}\ \bibnamefont {Peng}}, \bibinfo
  {author} {\bibfnamefont {Jinlong}\ \bibnamefont {Zhu}}, \bibinfo {author}
  {\bibfnamefont {Yijin}\ \bibnamefont {Zhang}}, \bibinfo {author}
  {\bibfnamefont {Xuliang}\ \bibnamefont {Chen}}, \bibinfo {author}
  {\bibfnamefont {Zhaorong}\ \bibnamefont {Yang}}, \bibinfo {author}
  {\bibfnamefont {Xiaodi}\ \bibnamefont {Liu}}, \bibinfo {author}
  {\bibfnamefont {Yanming}\ \bibnamefont {Ma}}, \bibinfo {author}
  {\bibfnamefont {Yusheng}\ \bibnamefont {Zhao}}, \bibinfo {author}
  {\bibfnamefont {Tomoko}\ \bibnamefont {Kagayama}}, \ and\ \bibinfo {author}
  {\bibfnamefont {Yoshihiro}\ \bibnamefont {Iwasa}},\ }\bibfield  {title}
  {\enquote {\bibinfo {title} {Ultrahigh pressure superconductivity in
  molybdenum disulfide},}\ }\href@noop {} {\bibfield  {journal} {\bibinfo
  {journal} {arXiv:1503.05331}\ } (\bibinfo {year} {2015})}\BibitemShut
  {NoStop}%
\bibitem [{\citenamefont {Somoano}\ \emph {et~al.}(1973)\citenamefont
  {Somoano}, \citenamefont {Hadek},\ and\ \citenamefont {Rembaum}}]{Somoano}%
  \BibitemOpen
  \bibfield  {author} {\bibinfo {author} {\bibfnamefont {R.~B.}\ \bibnamefont
  {Somoano}}, \bibinfo {author} {\bibfnamefont {V.}~\bibnamefont {Hadek}}, \
  and\ \bibinfo {author} {\bibfnamefont {A.}~\bibnamefont {Rembaum}},\
  }\bibfield  {title} {\enquote {\bibinfo {title} {Alkali metal intercalates of
  molybdenum disulfide},}\ }\href@noop {} {\bibfield  {journal} {\bibinfo
  {journal} {J. Chem. Phys.}\ }\textbf {\bibinfo {volume} {58}},\ \bibinfo
  {pages} {697--701} (\bibinfo {year} {1973})}\BibitemShut {NoStop}%
\bibitem [{\citenamefont {Somoano}\ \emph {et~al.}(1975)\citenamefont
  {Somoano}, \citenamefont {Hadek}, \citenamefont {Rembaum}, \citenamefont
  {Samson},\ and\ \citenamefont {Woollam}}]{Somoano2}%
  \BibitemOpen
  \bibfield  {author} {\bibinfo {author} {\bibfnamefont {R.~B.}\ \bibnamefont
  {Somoano}}, \bibinfo {author} {\bibfnamefont {V.}~\bibnamefont {Hadek}},
  \bibinfo {author} {\bibfnamefont {A.}~\bibnamefont {Rembaum}}, \bibinfo
  {author} {\bibfnamefont {S.}~\bibnamefont {Samson}}, \ and\ \bibinfo {author}
  {\bibfnamefont {J.~A.}\ \bibnamefont {Woollam}},\ }\bibfield  {title}
  {\enquote {\bibinfo {title} {The alkaline earth intercalates of molybdenum
  disulfide},}\ }\href@noop {} {\bibfield  {journal} {\bibinfo  {journal} {J.
  Chem. Phys.}\ }\textbf {\bibinfo {volume} {62}},\ \bibinfo {pages}
  {1068--1073} (\bibinfo {year} {1975})}\BibitemShut {NoStop}%
\bibitem [{Som()}]{Somoano3}%
  \BibitemOpen
  \bibfield  {title} {\enquote {\bibinfo {title} {Superconducting critical
  fields of alkali and alkaline-earth intercalates of {MoS$_2$}, author =
  {Woollam, John A. and Somoano, Robert B.}, journal = {Phys. Rev. B}, volume =
  {13}, pages = {3843--3853}, year = {1976}},}\ }\href@noop {} {\ }\BibitemShut
  {NoStop}%
\bibitem [{\citenamefont {Huang}\ \emph {et~al.}(2015)\citenamefont {Huang},
  \citenamefont {Xing},\ and\ \citenamefont {Xing}}]{GQHuang}%
  \BibitemOpen
  \bibfield  {author} {\bibinfo {author} {\bibfnamefont {G.~Q.}\ \bibnamefont
  {Huang}}, \bibinfo {author} {\bibfnamefont {Z.~W.}\ \bibnamefont {Xing}}, \
  and\ \bibinfo {author} {\bibfnamefont {D.~Y.}\ \bibnamefont {Xing}},\
  }\bibfield  {title} {\enquote {\bibinfo {title} {Prediction of
  superconductivity in li-intercalated bilayer phosphorene},}\ }\href@noop {}
  {\bibfield  {journal} {\bibinfo  {journal} {Appl. Phys. Lett.}\ }\textbf
  {\bibinfo {volume} {106}},\ \bibinfo {pages} {113107} (\bibinfo {year}
  {2015})}\BibitemShut {NoStop}%
\bibitem [{\citenamefont {Saito}\ \emph
  {et~al.}(2016{\natexlab{b}})\citenamefont {Saito}, \citenamefont {Nojima},\
  and\ \citenamefont {Iwasa}}]{YuSaito}%
  \BibitemOpen
  \bibfield  {author} {\bibinfo {author} {\bibfnamefont {Yu}~\bibnamefont
  {Saito}}, \bibinfo {author} {\bibfnamefont {Tsutomu}\ \bibnamefont {Nojima}},
  \ and\ \bibinfo {author} {\bibfnamefont {Yoshihiro}\ \bibnamefont {Iwasa}},\
  }\bibfield  {title} {\enquote {\bibinfo {title} {Gate-induced
  superconductivity in two-dimensional atomic crystals},}\ }\href@noop {}
  {\bibfield  {journal} {\bibinfo  {journal} {Supercond. Sci. Technol.}\
  }\textbf {\bibinfo {volume} {29}},\ \bibinfo {pages} {093001} (\bibinfo
  {year} {2016}{\natexlab{b}})}\BibitemShut {NoStop}%
\bibitem [{\citenamefont {Szcz{\c{e}}{\'s}niak}\ \emph
  {et~al.}(2014)\citenamefont {Szcz{\c{e}}{\'s}niak}, \citenamefont
  {Durajski},\ and\ \citenamefont {Szcz{\c{e}}{\'s}niak}}]{DominCLi}%
  \BibitemOpen
  \bibfield  {author} {\bibinfo {author} {\bibfnamefont {D}~\bibnamefont
  {Szcz{\c{e}}{\'s}niak}}, \bibinfo {author} {\bibfnamefont {A~P}\ \bibnamefont
  {Durajski}}, \ and\ \bibinfo {author} {\bibfnamefont {R}~\bibnamefont
  {Szcz{\c{e}}{\'s}niak}},\ }\bibfield  {title} {\enquote {\bibinfo {title}
  {Influence of lithium doping on the thermodynamic properties of graphene
  based superconductors},}\ }\href@noop {} {\bibfield  {journal} {\bibinfo
  {journal} {J. Phys.: Condens. Matter}\ }\textbf {\bibinfo {volume} {26}},\
  \bibinfo {pages} {255701} (\bibinfo {year} {2014})}\BibitemShut {NoStop}%
\bibitem [{\citenamefont {Ichinokura}\ \emph {et~al.}(2016)\citenamefont
  {Ichinokura}, \citenamefont {Sugawara}, \citenamefont {Takayama},
  \citenamefont {Takahashi},\ and\ \citenamefont {Hasegawa}}]{IchinokuraCaC1}%
  \BibitemOpen
  \bibfield  {author} {\bibinfo {author} {\bibfnamefont {Satoru}\ \bibnamefont
  {Ichinokura}}, \bibinfo {author} {\bibfnamefont {Katsuaki}\ \bibnamefont
  {Sugawara}}, \bibinfo {author} {\bibfnamefont {Akari}\ \bibnamefont
  {Takayama}}, \bibinfo {author} {\bibfnamefont {Takashi}\ \bibnamefont
  {Takahashi}}, \ and\ \bibinfo {author} {\bibfnamefont {Shuji}\ \bibnamefont
  {Hasegawa}},\ }\bibfield  {title} {\enquote {\bibinfo {title}
  {Superconducting calcium-intercalated bilayer graphene},}\ }\href@noop {}
  {\bibfield  {journal} {\bibinfo  {journal} {ACS Nano}\ }\textbf {\bibinfo
  {volume} {10}},\ \bibinfo {pages} {2761--2765} (\bibinfo {year}
  {2016})}\BibitemShut {NoStop}%
\bibitem [{\citenamefont {Chapman}\ \emph {et~al.}(2016)\citenamefont
  {Chapman}, \citenamefont {Su}, \citenamefont {Howard}, \citenamefont
  {Kundys}, \citenamefont {Grigorenko}, \citenamefont {Guinea}, \citenamefont
  {Geim}, \citenamefont {Grigorieva},\ and\ \citenamefont
  {Nair}}]{ChapmanCaC1}%
  \BibitemOpen
  \bibfield  {author} {\bibinfo {author} {\bibfnamefont {J.}~\bibnamefont
  {Chapman}}, \bibinfo {author} {\bibfnamefont {Y.}~\bibnamefont {Su}},
  \bibinfo {author} {\bibfnamefont {C.~A.}\ \bibnamefont {Howard}}, \bibinfo
  {author} {\bibfnamefont {D.}~\bibnamefont {Kundys}}, \bibinfo {author}
  {\bibfnamefont {A.~N.}\ \bibnamefont {Grigorenko}}, \bibinfo {author}
  {\bibfnamefont {F.}~\bibnamefont {Guinea}}, \bibinfo {author} {\bibfnamefont
  {A.~K.}\ \bibnamefont {Geim}}, \bibinfo {author} {\bibfnamefont {I.~V.}\
  \bibnamefont {Grigorieva}}, \ and\ \bibinfo {author} {\bibfnamefont {R.~R.}\
  \bibnamefont {Nair}},\ }\bibfield  {title} {\enquote {\bibinfo {title}
  {Superconductivity in {Ca}-doped graphene laminates},}\ }\href@noop {}
  {\bibfield  {journal} {\bibinfo  {journal} {Sci. Rep.}\ }\textbf {\bibinfo
  {volume} {6}},\ \bibinfo {pages} {23254} (\bibinfo {year}
  {2016})}\BibitemShut {NoStop}%
\bibitem [{\citenamefont {Giannozzi}\ \emph {et~al.}(2009)\citenamefont
  {Giannozzi}, \citenamefont {Baroni}, \citenamefont {Bonini}, \citenamefont
  {Calandra}, \citenamefont {Car}, \citenamefont {Cavazzoni}, \citenamefont
  {Ceresoli}, \citenamefont {Chiarotti}, \citenamefont {Cococcioni},
  \citenamefont {Dabo}, \citenamefont {Corso}, \citenamefont {de~Gironcoli},
  \citenamefont {Fabris}, \citenamefont {Fratesi},\ and\ \citenamefont
  {Gebauer}}]{Giannozzi2009A}%
  \BibitemOpen
  \bibfield  {author} {\bibinfo {author} {\bibfnamefont {P.}~\bibnamefont
  {Giannozzi}}, \bibinfo {author} {\bibfnamefont {S.}~\bibnamefont {Baroni}},
  \bibinfo {author} {\bibfnamefont {N.}~\bibnamefont {Bonini}}, \bibinfo
  {author} {\bibfnamefont {M.}~\bibnamefont {Calandra}}, \bibinfo {author}
  {\bibfnamefont {R.}~\bibnamefont {Car}}, \bibinfo {author} {\bibfnamefont
  {C.}~\bibnamefont {Cavazzoni}}, \bibinfo {author} {\bibfnamefont
  {D.}~\bibnamefont {Ceresoli}}, \bibinfo {author} {\bibfnamefont {G.~L.}\
  \bibnamefont {Chiarotti}}, \bibinfo {author} {\bibfnamefont {M.}~\bibnamefont
  {Cococcioni}}, \bibinfo {author} {\bibfnamefont {I.}~\bibnamefont {Dabo}},
  \bibinfo {author} {\bibfnamefont {A.~D.}\ \bibnamefont {Corso}}, \bibinfo
  {author} {\bibfnamefont {S.}~\bibnamefont {de~Gironcoli}}, \bibinfo {author}
  {\bibfnamefont {S.}~\bibnamefont {Fabris}}, \bibinfo {author} {\bibfnamefont
  {G.}~\bibnamefont {Fratesi}}, \ and\ \bibinfo {author} {\bibfnamefont
  {R.}~\bibnamefont {Gebauer}},\ }\bibfield  {title} {\enquote {\bibinfo
  {title} {{QUANTUM ESPRESSO}: a modular and open-source software project for
  quantum simulations of materials},}\ }\href@noop {} {\bibfield  {journal}
  {\bibinfo  {journal} {J. Phys. Condens. Matter}\ }\textbf {\bibinfo {volume}
  {21}},\ \bibinfo {pages} {395502} (\bibinfo {year} {2009})}\BibitemShut
  {NoStop}%
\bibitem [{\citenamefont {Giustino}(2017)}]{GiustinoEP}%
  \BibitemOpen
  \bibfield  {author} {\bibinfo {author} {\bibfnamefont {Feliciano}\
  \bibnamefont {Giustino}},\ }\bibfield  {title} {\enquote {\bibinfo {title}
  {Electron-phonon interactions from first principles},}\ }\href@noop {}
  {\bibfield  {journal} {\bibinfo  {journal} {Rev. Mod. Phys.}\ }\textbf
  {\bibinfo {volume} {89}},\ \bibinfo {pages} {015003} (\bibinfo {year}
  {2017})}\BibitemShut {NoStop}%
\bibitem [{\citenamefont {Giustino}(2014)}]{FelicianoGiustino}%
  \BibitemOpen
  \bibfield  {author} {\bibinfo {author} {\bibfnamefont {Feliciano}\
  \bibnamefont {Giustino}},\ }\href@noop {} {\emph {\bibinfo {title} {Materials
  modelling using density functional theory. Properties and predictions}}}\
  (\bibinfo  {publisher} {Oxford University Press, Oxford},\ \bibinfo {year}
  {2014})\BibitemShut {NoStop}%
\bibitem [{\citenamefont {Verble}\ and\ \citenamefont
  {Wieting}(1970)}]{vanderWalls}%
  \BibitemOpen
  \bibfield  {author} {\bibinfo {author} {\bibfnamefont {J.~L.}\ \bibnamefont
  {Verble}}\ and\ \bibinfo {author} {\bibfnamefont {T.~J.}\ \bibnamefont
  {Wieting}},\ }\bibfield  {title} {\enquote {\bibinfo {title} {Lattice mode
  degeneracy in {MoS$_2$} and other layer compounds},}\ }\href@noop {}
  {\bibfield  {journal} {\bibinfo  {journal} {Phys. Rev. Lett.}\ }\textbf
  {\bibinfo {volume} {25}},\ \bibinfo {pages} {362--365} (\bibinfo {year}
  {1970})}\BibitemShut {NoStop}%
\bibitem [{\citenamefont {Szcz{\c{e}}{\'s}niak}\ \emph
  {et~al.}(2017)\citenamefont {Szcz{\c{e}}{\'s}niak}, \citenamefont
  {Durajski},\ and\ \citenamefont {Jarosik}}]{SzczDurJar}%
  \BibitemOpen
  \bibfield  {author} {\bibinfo {author} {\bibfnamefont {R.}~\bibnamefont
  {Szcz{\c{e}}{\'s}niak}}, \bibinfo {author} {\bibfnamefont {A.P.}\
  \bibnamefont {Durajski}}, \ and\ \bibinfo {author} {\bibfnamefont {M.W.}\
  \bibnamefont {Jarosik}},\ }\bibfield  {title} {\enquote {\bibinfo {title}
  {Metallization and superconductivity in {Ca}-intercalated bilayer mos$_2$},}\
  } {\bibfield
  {journal} {\bibinfo  {journal} {J. Phys. Chem. Solids}\ }\textbf {\bibinfo
  {volume} {111}},\ \bibinfo {pages} {254 -- 257} (\bibinfo {year}
  {2017})}\BibitemShut {NoStop}%
\bibitem [{\citenamefont {Eliashberg}(1960)}]{Eliashberg1960A}%
  \BibitemOpen
  \bibfield  {author} {\bibinfo {author} {\bibfnamefont {G.~M.}\ \bibnamefont
  {Eliashberg}},\ }\bibfield  {title} {\enquote {\bibinfo {title} {Interactions
  between electrons and lattice vibrations in a superconductor},}\ }\href@noop
  {} {\bibfield  {journal} {\bibinfo  {journal} {Soviet Physics JETP}\ }\textbf
  {\bibinfo {volume} {11}},\ \bibinfo {pages} {696} (\bibinfo {year}
  {1960})}\BibitemShut {NoStop}%
\bibitem [{\citenamefont {Carbotte}(1990)}]{Carbotte1990A}%
  \BibitemOpen
  \bibfield  {author} {\bibinfo {author} {\bibfnamefont {J.~P.}\ \bibnamefont
  {Carbotte}},\ }\bibfield  {title} {\enquote {\bibinfo {title} {Properties of
  boson-exchange superconductors},}\ }\href@noop {} {\bibfield  {journal}
  {\bibinfo  {journal} {Rev. Mod. Phys.}\ }\textbf {\bibinfo {volume} {62}},\
  \bibinfo {pages} {1027} (\bibinfo {year} {1990})}\BibitemShut {NoStop}%
\bibitem [{\citenamefont {Szcz{\c{e}}{\'s}niak}(2006)}]{Szczesniak2006B}%
  \BibitemOpen
  \bibfield  {author} {\bibinfo {author} {\bibfnamefont {R.}~\bibnamefont
  {Szcz{\c{e}}{\'s}niak}},\ }\bibfield  {title} {\enquote {\bibinfo {title}
  {The numerical solution of the imaginary-axis {E}liashberg equations},}\
  }\href@noop {} {\bibfield  {journal} {\bibinfo  {journal} {Acta Phys. Pol.
  A}\ }\textbf {\bibinfo {volume} {109}},\ \bibinfo {pages} {179} (\bibinfo
  {year} {2006})}\BibitemShut {NoStop}%
\bibitem [{\citenamefont {Szcz{\c{e}}{\'s}niak}\ and\ \citenamefont
  {Durajski}(2016)}]{Szczesniak2016MgH6}%
  \BibitemOpen
  \bibfield  {author} {\bibinfo {author} {\bibfnamefont {R.}~\bibnamefont
  {Szcz{\c{e}}{\'s}niak}}\ and\ \bibinfo {author} {\bibfnamefont {A.~P.}\
  \bibnamefont {Durajski}},\ }\bibfield  {title} {\enquote {\bibinfo {title}
  {Superconductivity well above room temperature in compressed {MgH$_6$}},}\
  }\href@noop {} {\bibfield  {journal} {\bibinfo  {journal} {Front. Phys.}\
  }\textbf {\bibinfo {volume} {11}},\ \bibinfo {pages} {117406} (\bibinfo
  {year} {2016})}\BibitemShut {NoStop}%
\bibitem [{\citenamefont {Carbotte}\ and\ \citenamefont
  {Vashishta}(1970)}]{CARBOTTE1970227}%
  \BibitemOpen
  \bibfield  {author} {\bibinfo {author} {\bibfnamefont {J.P.}\ \bibnamefont
  {Carbotte}}\ and\ \bibinfo {author} {\bibfnamefont {P.}~\bibnamefont
  {Vashishta}},\ }\bibfield  {title} {\enquote {\bibinfo {title} {Condensation
  energy of a superconductor},}\ }\href@noop {} {\bibfield  {journal} {\bibinfo
   {journal} {Phys. Lett. A}\ }\textbf {\bibinfo {volume} {33}},\ \bibinfo
  {pages} {227 -- 228} (\bibinfo {year} {1970})}\BibitemShut {NoStop}%
\bibitem [{\citenamefont {S{\'o}lyom}(2011)}]{Solyom}%
  \BibitemOpen
  \bibfield  {author} {\bibinfo {author} {\bibfnamefont {J.}~\bibnamefont
  {S{\'o}lyom}},\ }\href@noop {} {\emph {\bibinfo {title} {Fundamentals of the
  Physics of Solids: Volume 3 - Normal, Broken-Symmetry, and Correlated
  Systems}}}\ (\bibinfo  {publisher} {Springer},\ \bibinfo {year}
  {2011})\BibitemShut {NoStop}%
\bibitem [{\citenamefont {Bardeen}\ \emph
  {et~al.}(1957{\natexlab{a}})\citenamefont {Bardeen}, \citenamefont {Cooper},\
  and\ \citenamefont {Schrieffer}}]{Bardeen1957A}%
  \BibitemOpen
  \bibfield  {author} {\bibinfo {author} {\bibfnamefont {J.}~\bibnamefont
  {Bardeen}}, \bibinfo {author} {\bibfnamefont {L.~N.}\ \bibnamefont {Cooper}},
  \ and\ \bibinfo {author} {\bibfnamefont {J.~R.}\ \bibnamefont {Schrieffer}},\
  }\bibfield  {title} {\enquote {\bibinfo {title} {Microscopic theory of
  superconductivity},}\ }\href@noop {} {\bibfield  {journal} {\bibinfo
  {journal} {Phys. Rev.}\ }\textbf {\bibinfo {volume} {106}},\ \bibinfo {pages}
  {162} (\bibinfo {year} {1957}{\natexlab{a}})}\BibitemShut {NoStop}%
\bibitem [{\citenamefont {Bardeen}\ \emph
  {et~al.}(1957{\natexlab{b}})\citenamefont {Bardeen}, \citenamefont {Cooper},\
  and\ \citenamefont {Schrieffer}}]{Bardeen1957B}%
  \BibitemOpen
  \bibfield  {author} {\bibinfo {author} {\bibfnamefont {J.}~\bibnamefont
  {Bardeen}}, \bibinfo {author} {\bibfnamefont {L.~N.}\ \bibnamefont {Cooper}},
  \ and\ \bibinfo {author} {\bibfnamefont {J.~R.}\ \bibnamefont {Schrieffer}},\
  }\bibfield  {title} {\enquote {\bibinfo {title} {Theory of
  superconductivity},}\ }\href@noop {} {\bibfield  {journal} {\bibinfo
  {journal} {Phys. Rev.}\ }\textbf {\bibinfo {volume} {108}},\ \bibinfo {pages}
  {1175} (\bibinfo {year} {1957}{\natexlab{b}})}\BibitemShut {NoStop}%
\end{thebibliography}%
%%%%%%%%%%%%%%%%%%%%%
%
\end{document}